\documentclass[pra,aps,preprint,superscriptaddress,12pt]{revtex4-2}

\usepackage{physics}
\usepackage{xspace}
\usepackage{graphicx}
\usepackage{setspace}

\usepackage[colorlinks, citecolor={blue}, linkcolor={black}, urlcolor={blue}]{hyperref}

\usepackage{titlesec}
\titleformat{\section}{\normalfont\Large\bfseries}{\thesection}{1ex}{}
\titleformat{\subsection}{\normalfont\large\bfseries}{\thesubsection}{1ex}{}
\renewcommand{\thesection}{\arabic{section}}
\renewcommand{\thesubsection}{\arabic{section}.\arabic{subsection}}

\makeatletter
\def\l@subsubsection#1#2{}
\makeatother

\begin{document}

\setstretch{1}

\title{Light-controlled van der Waals tunnel junctions:\texorpdfstring{\\}{} mechanisms, architectures, functionalities, and opportunities}

\author{Mohamed Shehabeldin}
\affiliation{Department of Physics, Boston College, Chestnut Hill, MA, USA}

\author{Xuguo Zhou}
\affiliation{Department of Physics, Boston College, Chestnut Hill, MA, USA}

\author{Ran Li}
\affiliation{Department of Materials Science and Engineering, Texas A$\&$M University, College Station, TX, USA}

\affiliation{Department of Electrical and Computer Engineering, Texas A$\&$M University, College Station, TX, USA}
\author{Pablo Jarillo-Herrero}
\affiliation{Department of Physics, Massachusetts Institute of Technology, Cambridge, MA, USA}

\author{Yuxuan Cosmi Lin}
\affiliation{Department of Materials Science and Engineering, Texas A$\&$M University, College Station, TX, USA}

\affiliation{Department of Electrical and Computer Engineering, Texas A$\&$M University, College Station, TX, USA}

\author{Jian Tang}
\email{Corresponding author: tanggv@bc.edu}
\affiliation{Department of Physics, Boston College, Chestnut Hill, MA, USA}

\author{Qiong Ma}
\email{Corresponding author: maqa@bc.edu}
\affiliation{Department of Physics, Boston College, Chestnut Hill, MA, USA}
\affiliation{The Schiller Institute for Integrated Science and Society, Boston College, Chestnut Hill, MA, USA}

\maketitle
\tableofcontents

\newpage
\section*{Abstract}

\textbf{
The phenomenon of electron tunneling has long been central to quantum transport and continues to provide a powerful framework for understanding and controlling electronic processes in solids. When combined with optical excitation, tunneling becomes a particularly rich platform for experiments, because light can drive nonequilibrium carrier populations and open transport pathways that are inaccessible without optical excitation. The emergence of van der Waals (vdW) materials has greatly expanded this opportunity by enabling atomically thin heterostructures with clean interfaces, engineered barriers, and highly tunable band alignment. In this review, we discuss the fundamental mechanisms of photo-assisted transport and the realization of vdW tunnel junctions, and show how they provide electrical access to nonequilibrium dynamics and collective excitations in quantum materials. We further examine emerging functionalities including photodetection, tunneling-driven light emission, sensing, and memory. Finally, we present a forward-looking perspective on new opportunities such as quantum-geometric probes, twist-resolved spectroscopy, moiré ferroelectricity, and scalable architectures for computing and sensing.
}

\section{Introduction}

Controlling electronic transport with external fields is a central goal of modern condensed-matter physics and device engineering. Whereas static electric fields regulate carrier motion through band bending and gating, time-dependent electromagnetic fields (light) provide a qualitatively different handle: they exchange discrete energy quanta with carriers, drive nonequilibrium populations, and open transport pathways that are unavailable otherwise. Tunnel junctions are especially sensitive to such driving because transmission through a barrier depends exponentially on carrier energy, so even weak photoexcitation can strongly modify current flow. Therefore, tunnel junctions under light can function simultaneously as probes of electronic excitations and as active optoelectronic elements~\cite{electricOptical_tunneling_review, ang2021physics, ahmad2023progress}.\\

In the classical picture, an electron is treated as a point particle encountering a hard-wall potential barrier: it can traverse a barrier of height $V_0$ only when its energy $E > V_0$, whereas transmission is strictly forbidden for $E < V_0$, leading to vanishing conductance through an insulating barrier. Quantum mechanics fundamentally revises this view by endowing particles with wave-like character, allowing the wavefunction to penetrate otherwise classically forbidden regions. For a finite barrier with $E < V_0$, the wavefunction decays exponentially as $\psi(x) \propto e^{-\kappa x}$, where 
$\kappa = \sqrt{2m^{*}(V_0 - E)}/\hbar$, $m^*$ is the effective carrier mass in the barrier, and $x$ is the coordinate into the barrier, resulting in a finite transmission probability $T \propto e^{-2\kappa d}$, with $d$ the barrier thickness~\cite{Sakurai2020}. Tunneling enables charge and spin transport across nanometer-scale insulating layers and forms the physical basis of tunnel junction-based electronic and spintronic devices~\cite{Datta_mesoscopic, tmr1}.

\subsection{Electrically- versus optically-controlled tunneling}

\begin{figure}
\centering
\includegraphics[width=6.5in]{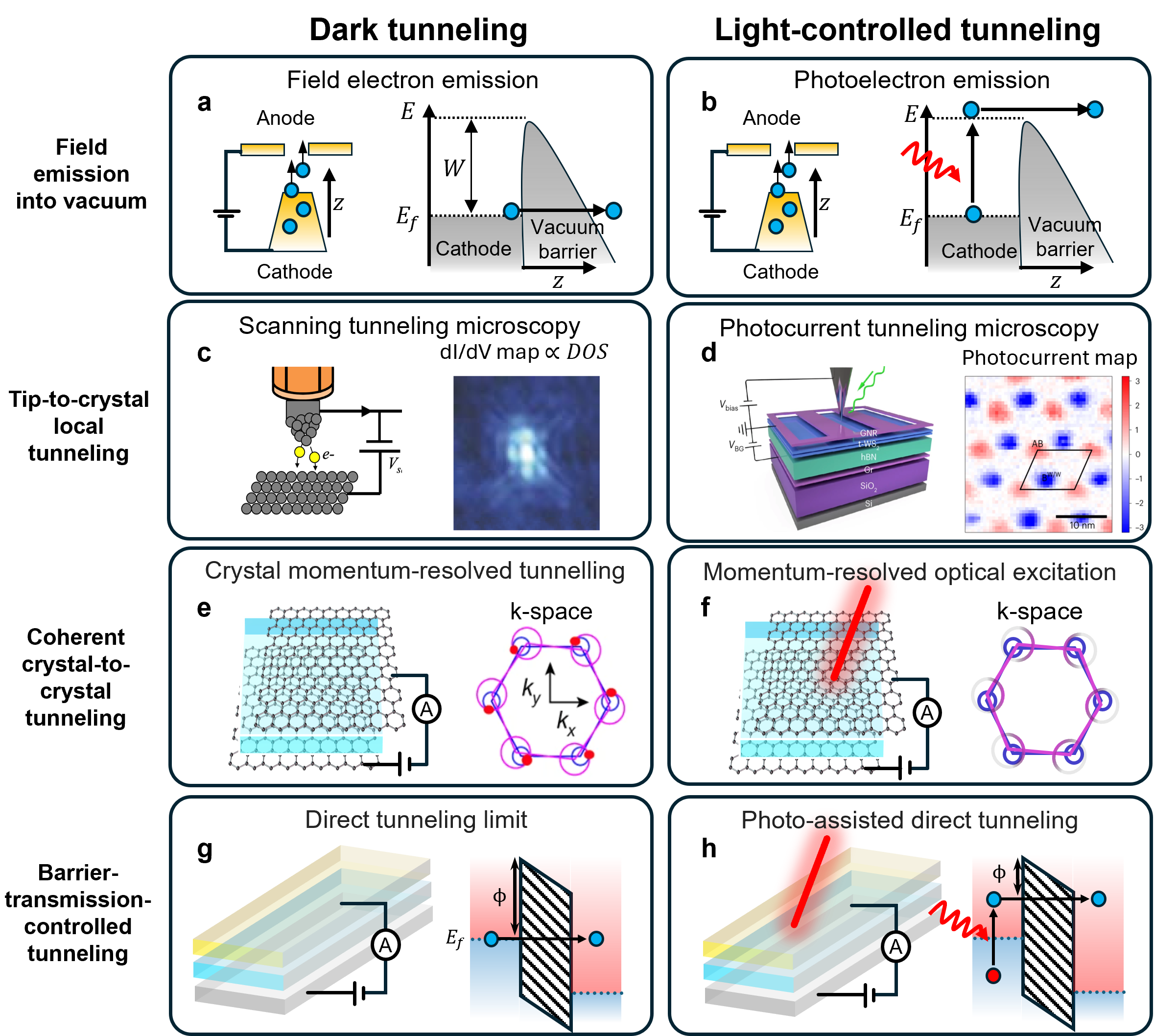}
\caption{
\textbf{Types of dark tunneling and their light-controlled counterparts.}
Left column: electrically driven or dark transport; right column: corresponding light-assisted cases.
\textbf{(a,b)} Field emission into vacuum: dark field emission under a strong electric field and photoinduced electron emission.
\textbf{(c,d)} Tip-to-crystal local tunneling: STM/STS, where $dI/dV$ probes the local density of states, and photocurrent tunneling microscopy under illumination. Adapted from Refs.~\cite{STM_Fe_review,STM_thesis,STM_photocurrent}.
\textbf{(e,f)} Crystal-to-crystal coherent tunneling: momentum-selective interlayer tunneling and its optical reweighting in $k$ space. Adapted from Ref.~\cite{inbar2023quantum}.
\textbf{(g,h)} Barrier-transmission-controlled tunneling: dark direct tunneling and photo-assisted direct tunneling, including channels involving discrete energy exchange $\hbar\omega$.
}
\label{fig1}
\end{figure}

Field electron emission, the ejection of electrons from a metal surface under a strong external electric field, stands among the earliest experimental manifestations of quantum tunneling, formalized by Fowler and Nordheim in 1928~\cite{FN_emission}. In this regime, the electric field tilts the surface potential barrier into a triangular profile [Fig.~\ref{fig1}(a)], and electrons tunnel through the barrier rather than surmounting it, with a current governed primarily by barrier shape and field strength~\cite{FN_emission, FN_emission_reformation}. Photoexcitation raises the energy of charge carriers, modifies their 
distribution, and modifies the weight of inelastic processes, including phonon- and plasmon-assisted pathways [Fig.~\ref{fig1}(b)]~\cite{inelastic_emitter_2019, 
inelastic_blasmonic_emitter, photoelectric_field_emisson}. In energy space, electrical bias reshapes the potential landscape, whereas light injects carriers into higher-energy initial states, effectively broadening the energy window available for transmission.\\

Scanning tunneling microscopy (STM) provides a local tunneling geometry in which electrons traverse a vacuum gap between a metallic tip and a sample surface. In conventional STM operation, the tip--sample distance is on the order of a few \r{A}ngstr\"{o}ms and the bias voltage $V$ is typically tens to hundreds of millivolts. Since this bias window is small compared with the several-electron-volt vacuum barrier set by the tip and sample work functions, the electric-field-induced tilt of the barrier is weak and the tunneling barrier is approximated as rectangular. Since the tip is metallic, its density of states can be approximated as constant over the relevant bias window. Under these conditions and in the weak-coupling (tip-sample) limit, the tunneling current $I$ is proportional to the local density of states (LDOS) of the sample, and the differential conductance $dI/dV$ maps the LDOS at energy $E_F + eV$, where $V$ is the applied bias voltage, $E_F$ is the sample Fermi energy, and $e$ the elementary charge [Fig.~\ref{fig1}(c)]~\cite{STM_Fe_review, STM_thesis}. Optical excitation creates nonequilibrium carrier occupations in the sample, producing spatially resolved photocurrent signals that encode the local optoelectronic response at the nanoscale. In photocurrent tunneling microscopy of twisted bilayer WS$_2$, this approach resolved the internal charge distribution of moir\'{e} excitons with sub-nanometer resolution [Fig.~\ref{fig1}(d)]~\cite{STM_photocurrent}.\\

Device-based tunneling, particularly in van der Waals (vdW) heterostructures that are the focus of this review, offers clean and versatile material interfaces, tunable barrier compositions, and controllable symmetry. In these crystal-to-crystal junctions, wavefunction overlap introduces approximate conservation of in-plane momentum $k_{\parallel}$, making tunneling momentum- and symmetry-selective [Fig.~\ref{fig1}(e)]~\cite{zhang2003band, inbar2023quantum}. Optical excitation can redistribute the populated momentum states in one of the electrodes through selective excitation of electronic states~\cite{sipe2000second, TMDs_DiXiao} [Fig.~\ref{fig1}(f)], thereby changing which interlayer transitions are momentum-matched and can contribute to tunneling. When momentum selectivity is weak, the barrier profile instead governs transport, and photoexcitation enables additional transmission via discrete photon exchange $\hbar\omega$ [Fig.~\ref{fig1}(h)].\\

Beyond single-particle processes, strong electromagnetic fields can coherently dress electronic states, creating Floquet sidebands that may redistribute tunneling probability and enable coherent control of transport pathways~\cite{gr_hBN_gr_floquet_theory, strongfield_tunneling_review, Anomalous_Floquet_tunneling_theory}. Photon absorption may also proceed through many-body processes such as carrier--carrier scattering~\cite{Gr/BN/Gr_Ma} and exciton--exciton annihilation~\cite{MoS2/BN/Gr_EEA}. Light-controlled tunnel junctions thus provide access to carrier dynamics and nonequilibrium states that electrical driving alone cannot reach, establishing light-controlled tunneling as a qualitatively richer probe of quantum transport~\cite{ electricOptical_tunneling_review, ahmad2023progress, multiphoton_SIS}.\\

\subsection{Advantages of vdW tunnel junctions}

Conventional tunneling devices face persistent challenges in achieving atomic-scale control over barrier thickness and uniformity, suppressing interfacial disorder for maintaining tunneling coherence~\cite{tunneling_spectroscopy_book}. These difficulties are more severe in spin-, symmetry-, or correlation-driven systems, where disorder and interface reconstructions can obscure the very quantum effects one seeks to probe~\cite{tunneling_in_vdw_book, AFMTJ_review}.\\

VdW heterostructures provide a natural route to overcome these limitations. Their atomically flat, dangling-bond-free surfaces eliminate the interfacial reconstruction and disorder that limit conventional epitaxial junctions. Unlike epitaxial heterostructures, where lattice matching imposes strict constraints on which materials can be combined~\cite{koma1999van, zheleva1994epitaxial}, vdW layers can be assembled in arbitrary combinations without the requirement for lattice matching, enabling deterministic control of barrier width at the atomic level and precise engineering of band alignment across a wide range of material pairs~\cite{VdW1, VdW2, lattice_mismathc_and_vdw}.\\

Beyond overcoming these structural limitations, vdW heterostructures offer distinct physical advantages that expand what tunnel junctions can do. Because vdW materials host semiconducting, magnetic, superconducting, and topological phases, tunneling can probe spin, valley, topology, quantum geometry, and electronic correlations~\cite{2D_magnets_Burch, TMDs_Mak, gemotry_spotlight_M, tang2024dual, tang2025quantum}. The reduced dimensionality and strong light--matter coupling of vdW materials enable pronounced photoexcitation effects on tunneling~\cite{strong_light_vdw_coupling_rev}.\\

\section{Mechanisms of light-assisted transport in tunnel junctions}

A tunnel junction comprises three essential elements: an emitter that supplies carriers, a transport barrier that sets the energetic constraints for transmission, and a collector that receives the transmitted carriers [Fig.~\ref{fig2}(a,b)]. Optical excitation can perturb these components in different ways: (i) it raises carrier energies in the emitter, creating nonequilibrium populations that subsequently relax and thermalize on ultrafast timescales~\cite{Gr/BN/Gr_Ma, ultrafast_thermalization}; (ii) it populates states within the barrier, including intrinsic as well as impurity or defect levels, thereby introducing intermediate states that mediate transport~\cite{chandni2015evidence, chandni2016signatures, photon_phonon_assisted_tunneling}; and (iii) it modifies the junction electrostatics through photogating or photoinduced band bending~\cite{Gr/TaO/Gr_PG, ju2014photoinduced, WSe2/Ta2NiSe5}. These mechanisms produce characteristic signatures in the current--voltage characteristics, temperature dependence, and spectral response, allowing the dominant process to be experimentally identified.\\

Although mechanisms (ii) and (iii) are also of considerable interest, they involve qualitatively different physics. In mechanism (ii), optical excitation occurs within the barrier itself, for example when the photon energy exceeds the intrinsic bandgap of the barrier material or resonantly addresses a localized defect level. Transport then becomes photoconductive: photogenerated carriers created inside the barrier are collected by the electrodes, altering the conduction mechanism and enabling spectroscopic readout of the barrier electronic structure~\cite{chandni2015evidence, chandni2016signatures, photon_phonon_assisted_tunneling, Gr-hBN-Gr-UV}. In mechanism (iii), absorbed photons create a persistent charge distribution near the barrier interface, either by trapping photoexcited carriers in interfacial states or by generating a photovoltage that bends the bands at the junction, effectively gating the device optically. This photogating can produce very large, slow photoresponses because the trapped charge accumulates over many carrier lifetimes, acting as a persistent electrostatic perturbation rather than a direct transport channel~\cite{Gr/TaO/Gr_PG, ju2014photoinduced, WSe2/Ta2NiSe5, MoS2/cPVP/AuNPs_Multibit}.\\

The primary focus of this review is mechanism (i), in which optical excitation raises carrier energies in the emitter and creates nonequilibrium populations that relax on ultrafast timescales. In this regime, the tunneling current serves as a sensitive probe of the emitter's electronic structure and carrier dynamics~\cite{Gr/BN/Gr_Ma, Gr/WSe2/Gr_tunneling_thermionic, Gr/BN/Gr_IR, photon_phonon_assisted_tunneling}. Owing to its intrinsic energy selectivity and femtosecond-to-picosecond temporal response, vertical tunneling grants access to physical properties that are difficult to resolve in conventional transport measurements, which are typically diffusive and governed by much slower relaxation processes~\cite{electricOptical_tunneling_review, tunneling_in_vdw_book, ma2023photocurrent}.\\

The following subsections delineate the transport regimes associated with mechanism (i), organized according to whether optical excitation predominantly enhances sub-barrier transmission probabilities or activates over-the-barrier carrier transport (thermionic emission and photoemission). These regimes differ in the photoinduced carrier properties within the emitter, specifically the degree of hot-carrier generation and the shape of the nonequilibrium distribution, which together determine how the effective barrier is traversed. Section~\ref{probes_of_materials} then reviews how photo-assisted transport in these regimes has been employed as a spectroscopic and dynamical probe of fundamental properties in emerging quantum materials.

\subsection{Photo-assisted tunneling: direct and Fowler--Nordheim regimes}

We start from the tunneling framework in the absence of light and then incorporate optical perturbations to the carrier distribution. The tunneling problem can be simplified using a free-particle picture within the Wentzel--Kramers--Brillouin (WKB) approximation, which captures the essential exponential dependence of the transmission on barrier parameters~\cite{Datta_mesoscopic, tunneling_spectroscopy_book, gr/BN/gr_no_ligh_science}.\\

\begin{figure}
\centering
\includegraphics[width=5.0in]{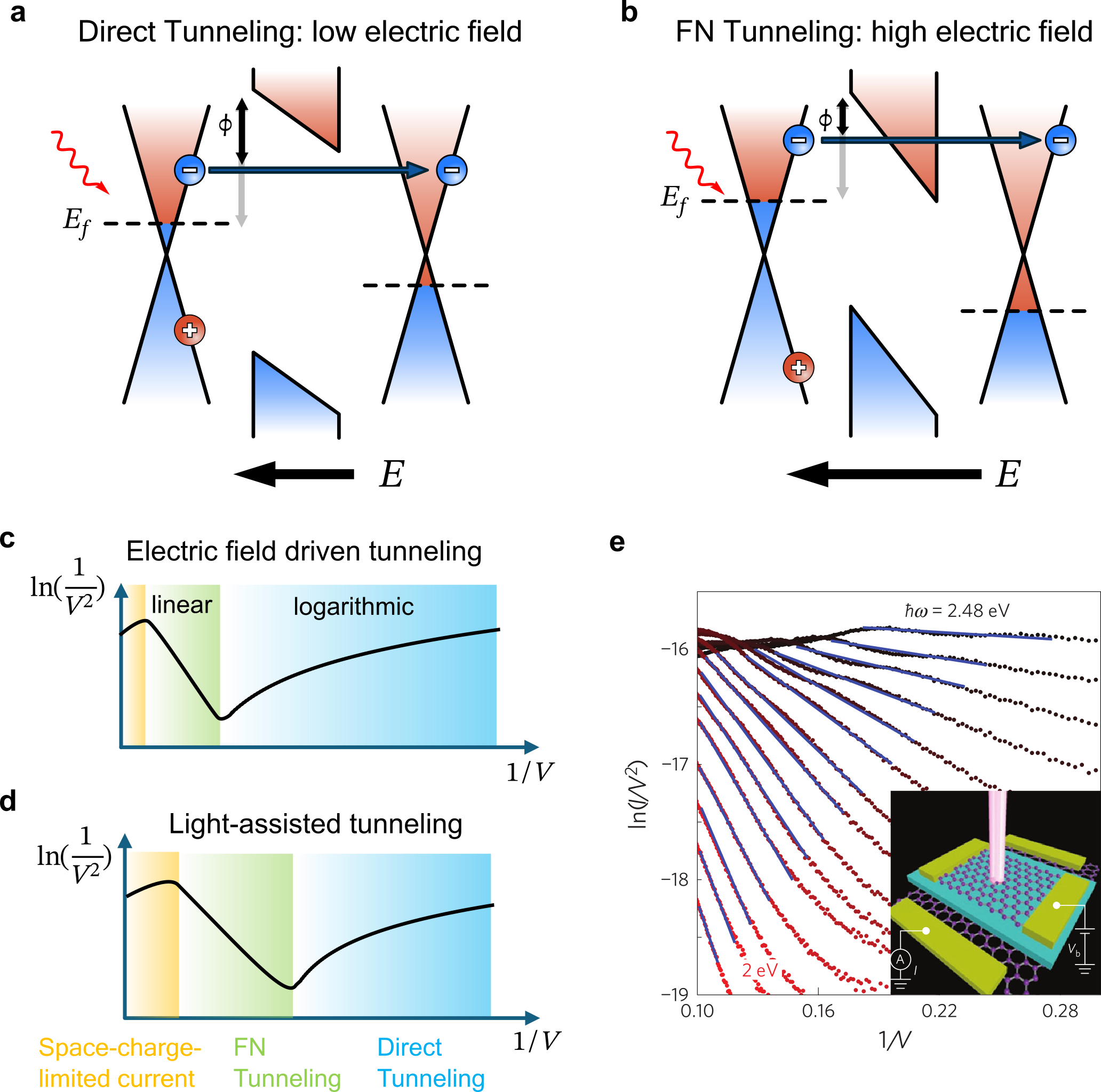}
\caption{
\textbf{Photo-assisted tunneling across direct-tunneling and Fowler--Nordheim regimes.}
Here, $E_F$ is the Fermi level, $\Phi$ the effective barrier height, and $I$, $V$ denote current and bias.
\textbf{(a)} Low-field direct tunneling through a relatively wide, weakly tilted barrier.
\textbf{(b)} High-field Fowler--Nordheim tunneling through a strongly tilted, triangular barrier.
\textbf{(c)} Representative dark FN plot, $\ln(I/V^2)$ versus $1/V$, showing three bias sectors from left to right: space-charge-limited-current, FN-tunneling, and direct-tunneling regimes.
\textbf{(d)} Under illumination, the FN plot is reshaped, with shifted crossovers and modified slopes and intercepts, indicating photon-modified transmission.
\textbf{(e)} Experimental $\ln(I/V^2)$ versus $1/V$ families at different $\hbar\omega$, showing systematic optical modulation of FN-type characteristics in graphene/hBN/graphene junctions. Adapted from Ref.~\cite{Gr/BN/Gr_Ma}.
}
\label{fig2}
\end{figure}

Within the WKB approximation, at moderate bias, the barrier is only weakly tilted, so transmission is dominated by sub-barrier penetration through a trapezoidal-like potential profile [Fig.~\ref{fig2}(a)]. The tunneling current scales linearly with the applied bias voltage $V$ and decays exponentially with the barrier thickness $d$ and barrier height $\Phi$ as:

\begin{equation}
I_{\mathrm{DT}} \propto V \exp\!\left(
-\frac{2d}{\hbar}\sqrt{2 m^{*}\Phi}
\right),
\label{eq:direct_tunneling}
\end{equation}
where $m^{*}$ is the effective carrier mass in the barrier and $\hbar$ is the reduced Planck constant. The exponential factor directly reflects the WKB transmission through a rectangular potential barrier. This transport regime is referred to as direct tunneling.\\

At higher electric fields, the applied bias progressively tilts the barrier into a triangular profile, substantially enhancing transmission near the barrier top [Fig.~\ref{fig2}(b)]. This transport regime is referred to as Fowler--Nordheim (FN) tunneling, with current given by:

\begin{equation}
I_{\mathrm{FN}} \propto V^{2} \exp\!\left(
-\frac{4d}{3 e \hbar V}\sqrt{2 m^{*}\Phi^{3}}
\right),
\label{eq:FN_tunneling}
\end{equation}
where $e$ is the elementary charge. In contrast to direct tunneling [Eq.~\ref{eq:direct_tunneling}], FN tunneling is marked by a quadratic prefactor in $V$ and an exponential sensitivity to $1/V$, reflecting the triangular barrier shape under strong electric fields.\\

For experimental identification, Eq.~\ref{eq:FN_tunneling} can be recast in linearized form, 

\begin{equation}
\ln\!\left(\frac{I}{V^{2}}\right) = \ln\!\left(\frac{A}{\Phi}\right) - \frac{4\sqrt{2m^{*}}}{3\hbar e}\, \frac{{\Phi}^{3/2}}{V}, 
\end{equation} 

where $A$ is an effective prefactor that collects device-dependent factors such as the junction area and other slowly varying barrier parameters. A linear dependence of $\ln(I/V^{2})$ on $1/V$ therefore provides a characteristic signature of FN tunneling. The crossover between direct tunneling and FN tunneling thus provides a key benchmark for identifying the dominant transport mechanism in tunnel junctions [Fig.~\ref{fig2}(c)].\\

A primary mechanism by which light modulates the tunneling current is the promotion of carriers in the emitter into nonequilibrium high-energy states. These hot carriers therefore experience a reduced effective tunnel barrier, leading to an enhanced tunneling probability and an increase in current. A characteristic experimental signature of photo-assisted tunneling is that optical excitation shifts the onset of FN tunneling to substantially lower electric fields than in the dark, as photoexcited carriers effectively experience a thinner and lower triangular barrier [Fig.~\ref{fig2}(d)]. In the FN plot, photon-induced barrier lowering appears as a systematic decrease in slope, since the slope scales as $\Phi^{3/2}$, and a corresponding increase in intercept, since the intercept scales as $\ln(A/\Phi)$, enabling quantitative distinction from thermionic transport, which will be discussed in the following subsection.\\

This mechanism has been extensively studied in graphene/hexagonal boron nitride (hBN)/graphene heterostructures among others~\cite{Gr/BN/Gr_Ma, Gr/BN/Gr_Si3N4_waveguides, Gr/BN/Gr_IR}. Fig.~\ref{fig2}(e) shows measured families of $\ln(I/V^2)$ versus $1/V$ at different photon energies $\hbar\omega$: as $\hbar\omega$ increases, the slope decreases and the intercept rises, directly reflecting the progressive reduction of the effective barrier by photoexcitation~\cite{Gr/BN/Gr_Ma}.

\subsection{Thermionic and photoemission transport}
\label{thermionic_subsection}

In contrast to photo-assisted tunneling, where photoexcited carriers remain below the barrier maximum and traverse the junction via quantum tunneling, light can also enable charge transport by providing carriers with sufficient energy to \textit{classically} surmount the barrier \cite{MoS2/BN/Gr_UV, MoS2/BN/Gr_EEA, WSe2/BN/Gr_Xiadong_Auger, WSe2/BN/Gr_PhilipKim_Auger, Gr/BN/Gr_Ma,Gr/BN/Gr_IR}. In this regime, the barrier no longer acts as a tunneling obstacle but instead functions as an energy filter, and the photocurrent is governed by the population of carriers with energies exceeding the barrier height $\Phi$.\\

From a unifying distribution-function perspective, the total current can be expressed schematically as
\begin{equation}
I_{\mathrm{sub}} \;\propto\; \int_{E<\Phi} f(E)\,T(E)\,dE,
\end{equation}

\begin{equation}
I_{\mathrm{over}} \;\propto\; \int_{E\ge\Phi} f(E)\,dE,
\end{equation}
where $f(E)$ is the nonequilibrium carrier distribution. Whereas photon-assisted tunneling samples the sub-barrier portion of $f(E)$ weighted by the transmission probability $T(E)$, over-the-barrier transport is controlled by the high-energy tail of the distribution.\\

Over-the-barrier transport can arise through several pathways, depending on how carriers acquire sufficient energy to surmount the barrier: direct photoemission, many-body energy upconversion, photothermionic emission and bolometric responses.\\

\begin{figure}
\centering
\includegraphics[width=5in]{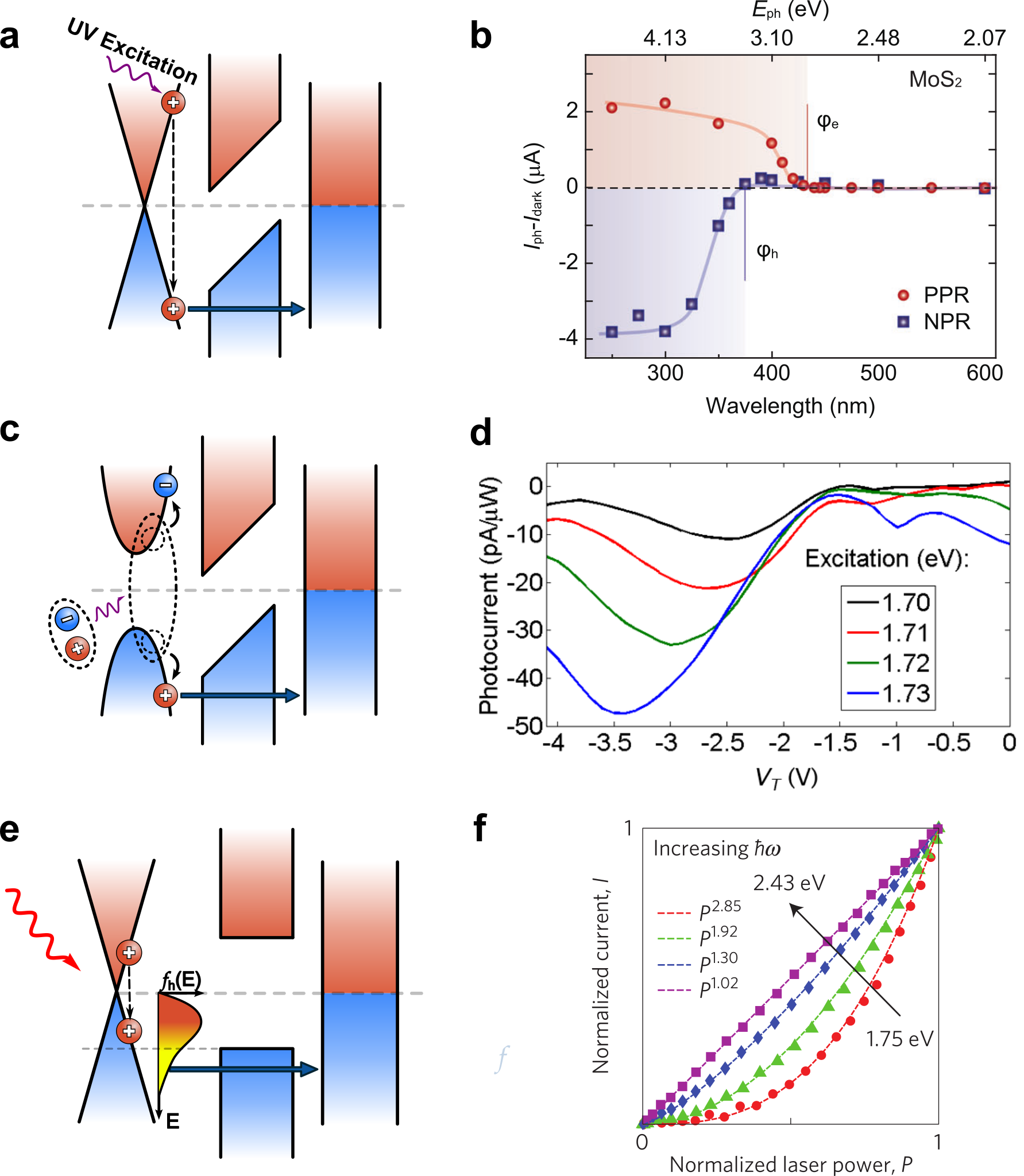}
\caption{
\textbf{Over-barrier hole-emission pathways in light-controlled vdW junctions.}
Here $I_{\mathrm{ph}}=I_{\mathrm{light}}-I_{\mathrm{dark}}$, $E_F$ is the Fermi-level reference, $E_{\mathrm{ph}}=\hbar\omega$ is photon energy, and $V_b$ (or $V_T$) is applied bias.
\textbf{(a,b)} Direct high-energy excitation:
(a) holes are photoexcited above the effective barrier and emitted over it;
(b) $(I_{\mathrm{ph}}-I_{\mathrm{dark}})$ versus wavelength shows threshold-like spectral behavior (PPR/NPR) consistent with barrier-defined internal photoemission. Adapted from Ref.~\cite{MoS2/BN/Gr_UV}.
\textbf{(c,d)} Many-body upconversion:
(c) exciton-mediated energy transfer (EEA/Auger-like) generates hot carriers that overcome the barrier;
(d) bias-dependent photocurrent exhibits excitonic resonances with enhanced current attributed to Auger-assisted channels. Adapted from Ref.~\cite{WSe2/BN/Gr_Xiadong_Auger}.
\textbf{(e,f)} Photothermal route:
(e) carrier heating broadens the distribution function of holes $f_h(E)$, and its high-energy tail contributes to over-barrier emission;
(f) photocurrent $I$ as a function of excitation laser power $P$, measured at constant interlayer bias, shows systematic power dependence consistent with thermally driven carrier redistribution. Adapted from Ref.~\cite{Gr/BN/Gr_Ma}.
}
\label{fig3}
\end{figure}

\textbf{(i) Direct internal photoemission.} When a photon excites an electron (or hole) directly into a state with energy exceeding the barrier height ($E > \Phi$), that carrier is emitted over the barrier without any prior energy redistribution among carriers [Fig.~\ref{fig3}(a)]. The hallmark of this mechanism is that the photocurrent onset is governed by the barrier height rather than the bandgap of the absorbing layer: although absorption still requires $\hbar\omega \geq E_g$, the relevant spectral threshold for photoemission is $\hbar\omega = \Phi$. Li \textit{et al.} demonstrated this in MoS$_2$/hBN/graphene junctions, where the electron and hole tunneling barriers of the hBN layer set two independent spectral thresholds at $\approx 2.8$~eV and $\approx 3.3$~eV, respectively, corresponding to positive and negative photoresponse cutoffs [Fig.~\ref{fig3}(b)]~\cite{MoS2/BN/Gr_UV}.\\

\textbf{(ii) Indirect excitation via many-body energy upconversion.} Beyond direct excitation, carriers can acquire sufficient energy to surmount the barrier through many-body interactions that concentrate the energy of multiple electron--hole pairs onto a single carrier. Photoexcitation creates electron--hole pairs; through subsequent interactions, one pair can recombine and transfer its energy to another carrier in the system rather than emitting it as a photon or phonon, promoting that carrier to a hot state that exceeds $\Phi$. Two representative examples of such energy upconversion are exciton--exciton annihilation (EEA) and Auger recombination. In exciton--exciton annihilation (EEA)~\cite{EEA_1, EEA_2}, two excited electron--hole pairs interact such that one recombines and transfers its energy to the other, generating a carrier with kinetic energy on the order of twice the optical gap. In Auger recombination~\cite{Auger_1, Auger_2}, the recombination energy of one electron--hole pair is instead transferred to a third carrier, electron or hole, promoting it to a hot state. In either case, the resulting energetic carrier can exceed the tunnel barrier height and contribute to over-barrier photocurrent, as illustrated schematically in Fig.~\ref{fig3}(c).\\

An example consistent with this picture is shown in Fig.~\ref{fig3}(d), where bias-dependent photocurrent traces measured at different excitation energies exhibit pronounced resonant enhancement at the charged-exciton transition energy~\cite{WSe2/BN/Gr_Xiadong_Auger}. This behavior indicates that the over-barrier photocurrent is strongly influenced by excitonic populations and interaction-driven energy redistribution, rather than by single-particle absorption alone. More broadly, these many-body upconversion processes show that over-barrier photocurrent can probe nonequilibrium high-energy carriers generated through collective relaxation pathways, in contrast to the thermally broadened distributions discussed below for photothermionic and bolometric responses.\\

\textbf{(iii) Photothermionic and bolometric responses.} A related class of photoresponses arises when optical absorption thermalizes the photoexcited carriers into a hot Fermi--Dirac distribution characterized by an elevated electronic temperature $T_e$. Within the distribution-function picture introduced above, photoexcitation still reshapes $f(E)$, but now through thermalization rather than through direct excitation or many-body upconversion. The key question is which portion of the heated distribution dominates transport: the high-energy tail above $\Phi$, or the sub-barrier weight where the tunneling probability is modified.\\

In the photothermionic pathway, optical absorption elevates the electronic temperature $T_e$, broadening the Fermi--Dirac distribution such that a finite fraction of carriers occupies states above the barrier $\Phi$ and surmounts it [Fig.~\ref{fig3}(e)]~\cite{photothermionic_1, photothermionic_2}. This process is ultrafast, as carrier thermalization typically occurs on timescales from tens of femtoseconds to picoseconds~\cite{Gr/BN/Gr_Ma, ultrafast_thermalization, Gr/WSe2/Gr_tunneling_thermionic}. A key experimental diagnostic is a superlinear dependence of the photocurrent on optical power: while $T_e$ rises only sublinearly with laser power (e.g., $T_e \sim P^{1/3}$ in graphene), the population of carriers above the barrier grows exponentially as $\propto \exp[-\Phi/(k_{\mathrm{B}}T_e)]$, leading to an overall superlinear photoresponse~\cite{Gr/WSe2/Gr_tunneling_thermionic, Gr/BN/Gr_Ma}. This thermionic regime is most prominent when the barrier remains relatively high compared with the excitation conditions. As the applied bias or photon energy increases, the effective barrier is reduced and the photocurrent crosses over toward a nearly linear power dependence, consistent with increasing contribution from direct photo-assisted tunneling. In the graphene/hBN/graphene device of Ref.~\cite{Gr/BN/Gr_Ma}, this crossover is seen directly in the power-law exponent changing from strongly superlinear at low photon energy to nearly linear at higher photon energy [Fig.~\ref{fig3}(f)].\\

Photoinduced heating can also generate a bolometric contribution to sub-barrier transport, alongside over-barrier emission. When the barrier is not too high, thermal broadening of the carrier distribution can substantially alter the tunneling current without requiring carriers to surmount the barrier. For example, in graphene/hBN/graphene junctions where transport proceeds through localized defect states in the hBN barrier, radiation-induced electron heating broadens the electronic distribution function $f(E)$ around the defect energy and changes the energy-averaged tunneling probability. At finite bias, this thermal contribution appears as a bolometric response across the tunnel junction, and the photoinduced current $I$ can be approximated as $I \approx I_{\mathrm{dark}}(T_0+\delta T)-I_{\mathrm{dark}}(T_0) \approx (dI_{\mathrm{dark}}/dT)\,\delta T$~\cite{Gr/BN/Gr_IR}, where $I_{\mathrm{dark}}(T)$ is the dark tunneling current at temperature $T$, $T_0$ is the base temperature and $\delta T$ is the light-induced electronic temperature rise. The signal therefore reflects the temperature dependence of the tunneling conductance. Experimentally, the photocurrent was found to be proportional to $d^2 I_{\mathrm{dark}}/dV^2$, where $V$ is the applied bias voltage. This curvature is related to $dI_{\mathrm{dark}}/dT$, supporting a bolometric interpretation of the finite-bias response~\cite{Gr/BN/Gr_IR}. At zero bias, asymmetric heating of the two graphene electrodes can additionally generate a Seebeck-type current across the barrier, contributing to the photoresponse even in the absence of an applied field~\cite{Gr/BN/Gr_IR}.

\section{Tunnel junctions as probes of materials}
\label{probes_of_materials}

Unlike conventional lateral transport, which reflects carrier motion within a material and is often dominated by scattering and diffusion, a tunnel junction probes the quantum transfer of carriers across an energy barrier. Because the transmission probability depends exponentially on carrier energy and wavefunction overlap, tunneling is sensitive to electronic structure and nonequilibrium carrier populations. Under optical excitation, changes in carrier distribution, many-body interactions, and collective excitations are converted directly into electrical signals, allowing the junction to function as an energy- and state-selective electrical readout of microscopic dynamics. VdW heterostructures are particularly suited for this purpose: atomically sharp interfaces permit the barrier thickness, composition, and band alignment to be engineered with near-atomic precision while preserving intrinsic material properties. By appropriate junction design, tunneling can therefore probe charge and energy relaxation, excitonic processes, phonon coupling, spin, polarization, and momentum selectivity. We begin by examining how the tunneling barrier is realized and how junction architecture determines the physical processes accessed.

\subsection{Barrier realization and junction architecture}

Tunnel junctions can be realized in a variety of structures with different choices of barrier and electrode materials. Here we focus on how the tunneling barrier is formed. Broadly, two configurations exist: the barrier can be created by inserting a dielectric spacer layer between electrodes, or it can arise intrinsically from band offsets at a semiconductor heterointerface. In the former, the barrier height, thickness, and defect landscape are directly controlled through the choice of spacer material and thickness. In the latter, the barrier is set by the band alignment at the interface, which determines which carrier species participate in transport. The manner in which the barrier is realized therefore strongly influences the physical information encoded in the tunneling current, including carrier energy distribution, charge accumulation, excitonic correlations, and the degree of spin or momentum selectivity. We accordingly classify junctions into these two broad architectures and discuss them separately below.

\subsubsection{Spacer-defined tunnel junctions}

In this architecture, an atomically thin dielectric or semiconducting spacer is inserted between the emitter and collector to create a well-defined tunneling barrier. The barrier height, thickness, and defect landscape can be systematically engineered through control of the spacer thickness, material composition, and defect density.\\

A canonical example among vdW junctions is the graphene/hBN/graphene vertical junction, where hBN serves as a robust and atomically uniform tunnel barrier~\cite{gr/BN/gr_no_ligh_science, resonant_tunneling_2d, Gr_resonant2}. Under optical excitation, this platform enables hot-carrier extraction~\cite{Gr/BN/Gr_Ma}, infrared photoresponse with strong electrostatic tunability~\cite{Gr/BN/Gr_IR}, and waveguide-integrated operation~\cite{Gr/BN/Gr_Si3N4_waveguides}. In thicker-barrier configurations, Fowler--Nordheim tunneling of 
carriers from the graphene electrodes into the hBN band edges 
enables deep-ultraviolet electroluminescence~\cite{Gr-hBN-Gr-UV}.\\

Replacing one graphene electrode with a transition metal dichalcogenide (TMD) extends this platform to excitonic photoresponse pathways. Representative phenomena include electrically detected two-photon absorption in WS$_2$~\cite{WS2/BN/Gr_bifunctional} and photocurrent generation governed by exciton--exciton annihilation and Auger-assisted upconversion in MoS$_2$/hBN/graphene and WSe$_2$/hBN/graphene~\cite{MoS2/BN/Gr_EEA, WSe2/BN/Gr_Xiadong_Auger, WSe2/BN/Gr_PhilipKim_Auger}. The same barrier-engineered design also produces a barrier-defined deep-ultraviolet cutoff in MoS$_2$/hBN/graphene~\cite{MoS2/BN/Gr_UV}. Replacing hBN with a semiconducting spacer such as WSe$_2$ (graphene/WSe$_2$/graphene) enables gate-tunable, energy-filtered photocarrier collection through the photothermionic effect~\cite{Gr/WSe2/Gr_tunneling_thermionic}.\\

Spacer-defined architectures further support spintronic and memory functionalities, including magnetic-contact devices based on WSe$_2$/hBN/Fe$_3$GeTe$_2$ (FGT), chiral photoexcitation platforms based on InSe/hBN/graphene [Fig.~\ref{fig4}(a,b)]~\cite{ WSe2/BN/FGT_EL, InSe/BN/Gr_circular}, and floating-gate hybrids where photo-assisted tunneling directly couples to nonvolatile storage in PtS$_2$/hBN/graphene~\cite{PtS2/BN/Gr}.

\subsubsection{Band-alignment-defined tunnel junctions}

\begin{figure}
\centering
\includegraphics[width=6.5in]{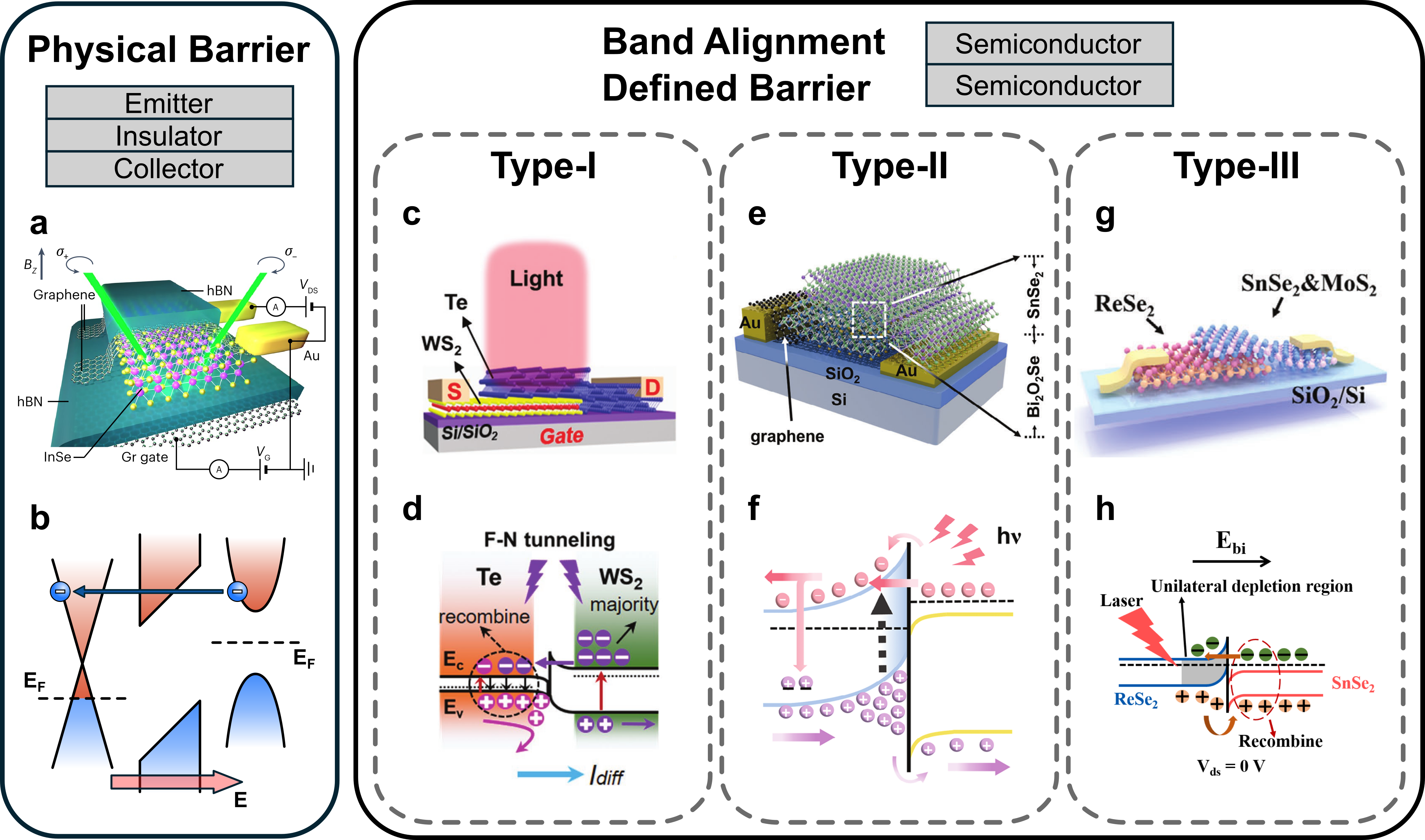}
\caption{\textbf{Device realizations of tunneling in 2D heterostructures: physical versus interface-defined barriers.}
\textbf{(a,b)} Physical-barrier device (InSe/hBN/graphene): optical spin/chirality-selective excitation in InSe is filtered by the hBN tunnel barrier and collected by graphene. Adapted from Ref.~\cite{InSe/BN/Gr_circular}.
\textbf{(c--h)} Band-alignment-defined tunnel junction devices. Three classes are shown, classified by the relative positions of the conduction- and valence-band edges.
\textbf{(c,d)} Type-I (straddling alignment): illustrated here with a WS$_2$/Te heterostructure, in which illumination drives carrier-selective interfacial transfer and tunneling-dominant extraction. Adapted from Ref.~\cite{WS2/Te}.
\textbf{(e,f)} Type-II (staggered alignment): illustrated here with a Bi$_2$O$_2$Se/SnSe$_2$ heterostructure, where built-in field together with gating or bias tunes interfacial transfer, gain, and response polarity. Adapted from Ref.~\cite{Bi2O2Se/SnSe2}.
\textbf{(g,h)} Type-III (broken-gap alignment): illustrated here with a ReSe$_2$/SnSe$_2$ heterostructure, where unilateral depletion enables strong zero- or low-bias photoresponse. Adapted from Ref.~\cite{ReSe2/SnSe2}.
These platforms show that light-controlled tunneling can be realized either with an explicit insulating barrier or through interfacial band alignment and electrostatics.
}
\label{fig4}
\end{figure}

In this architecture, the tunneling barrier arises intrinsically from band offsets at a semiconductor heterointerface, without any intervening dielectric spacer. The barrier height and profile are therefore set by the relative positions of the conduction and valence band edges of the two materials, in addition to interfacial band bending, which can generate a sharp and narrow interfacial potential~\cite{Zhang2012BandBending, Anderson1960}. Depending on this alignment, semiconductor heterojunctions are classified as type-I (straddling), type-II (staggered), or type-III (broken-gap), as shown in Fig.~\ref{fig4}~\cite{heterojunctions_1, heterojunctions_2}. This alignment determines which carrier species dominate interlayer transport, how efficiently photocarriers are separated, and whether transport proceeds via thermionic emission, band-to-band tunneling, or resonant injection pathways~\cite{ahmad2023progress}. We discuss these three cases in turn.\\

In a type-I heterojunction, the bandgap of one semiconductor is entirely contained within that of the other [Fig.~\ref{fig4}(d)], so both electrons and holes are confined to the same layer. This suppresses minority carrier accumulation at the interface and reduces interfacial recombination, relaxing the conventional trade-off between gain and recombination losses. For example, WS$_2$/Te heterostructures have been engineered such that photo-induced tunneling preferentially transfers majority electrons across the interface while suppressing minority carriers [Fig.~\ref{fig4}(c,d)], thereby reducing recombination and enabling high-responsivity imaging operation~\cite{WS2/Te}. Another example is WS$_2$/PtS$_2$, where the type-I alignment creates a transmission barrier that suppresses leakage current and enhances rectification, yielding low-noise photodetection governed by tunneling injection rather than diffusive transport~\cite{WS2/PtS2}.\\

In type-II (staggered) heterojunctions, the band offsets are arranged such that the conduction band minimum and valence band maximum reside in different materials [Fig.~\ref{fig4}(f)]. This staggered alignment enables field-controlled interlayer carrier transfer while maintaining strong electrostatic tunability. In Bi$_2$O$_2$Se/SnSe$_2$ heterostructures, potential engineering creates a deep hole trapping potential while narrowing the electron barrier under reverse bias, enhancing gain while preserving microsecond-scale response [Fig.~\ref{fig4}(e,f)]~\cite{Bi2O2Se/SnSe2}. In many vdW stacks, the interfacial alignment can be continuously tuned by gate fields, enabling controlled transitions between distinct transport regimes. PdSe$_2$/MoTe$_2$ can be electrostatically reconfigured between staggered and broken-gap alignment to switch photoresponse polarity and magnitude, bridging sensing and analog processing functionalities~\cite{PdSe2/MoTe2_processing}. Dual-gated MoS$_2$/WSe$_2$ devices similarly transition between rectifying and Esaki-diode behavior as the band alignment is tuned; in the Esaki regime, the junction supports interband tunneling between the conduction and valence bands of the two materials with negative differential resistance~\cite{MoS2/WSe2_TunnelDiode_gated}.\\

Type-III (broken-gap) junctions represent the extreme alignment case, where the valence-band maximum of one material lies above the conduction-band minimum of the other [Fig.~\ref{fig4}(h)]. Charge redistribution at the interface creates a triangular potential barrier that carriers traverse by direct tunneling at low bias and Fowler-Nordheim tunneling at higher bias. Such alignments support ultralow dark current, large photocurrent on/off ratios, and high-speed photoresponse~\cite{SnSe2/MoTe2}. A PtS$_2$/WSe$_2$ broken-gap heterostructure has leveraged this architecture to achieve a rectification ratio approaching $10^8$ and microsecond response, illustrating how band-to-band tunneling reduces transit-time limitations relative to diffusive carrier collection~\cite{PtS2/WSe2}. Broken-gap junctions can also operate at zero applied bias and directly encode intrinsic material properties: in ReSe$_2$/SnSe$_2$ devices [Fig.~\ref{fig4}(g,h)], photoinduced tunneling generates a measurable zero-bias photocurrent, while the strong in-plane anisotropy and linear dichroism of ReSe$_2$ enable polarization-sensitive detection without external polarizers~\cite{ReSe2/SnSe2}. Gate control further allows access to Esaki and backward-diode operation, enriching the nonlinear transport landscape accessible to photo-assisted tunneling studies~\cite{SnSe2/MoTe2, Bi2Se3/MoTe2}.\\

A closely related architecture is the Schottky junction, where one electrode is metallic and the barrier is set by the alignment between the metal Fermi level and the semiconductor band edge. In vdW geometries, graphene serves naturally as the metallic contact, and its low density of states makes the Schottky barrier strongly tunable by gate voltage. In graphene/WSe$_2$/graphene junctions, this tunability enables energy-filtered photocarrier collection through the photothermionic effect~\cite{Gr/WSe2/Gr_tunneling_thermionic}.\\

With this structural perspective established, we now turn to the microscopic physics through which optical excitation modifies tunneling transport. We begin with charge and energy dynamics, where hot-carrier generation and energy filtering define the spectral response and govern the competition between tunneling and thermalization. We then discuss phonon-mediated processes, including inelastic tunneling, hot-phonon bottlenecks, and bolometric pathways that emerge when vibrational relaxation dominates. Next, we address excitonic physics, where exciton dissociation, interlayer excitons, and many-body recombination processes are converted into tunneling currents. Finally, we examine spin and magnetic degrees of freedom, where angular-momentum transfer and magnetic selectivity enable spin- and valley-polarized tunneling and establish routes toward opto-spintronics in vdW junctions.

\subsection{Ultrafast charge dynamics}

Unlike lateral transport, which is governed by scattering, disorder, and diffusion over micrometer-scale paths, vertical tunneling transfers carriers through a nanometer-thick barrier via wavefunction overlap, without any classical propagation through the material. Vertical tunneling is therefore an intrinsically fast and clean conduction pathway. Furthermore, in a tunnel device, the junction itself serves as the primary element controlling the current. Even when measured under steady-state conditions, the current directly reflects microscopic processes occurring at the barrier on femtosecond to picosecond timescales, allowing steady-state measurements to encode information about ultrafast carrier dynamics and to compete with intrinsic carrier--carrier scattering processes within the emitter.\\

A direct demonstration of this principle was reported in graphene/hBN/graphene vertical junctions by Ma et al.~\cite{ Gr/BN/Gr_Ma}, who measured the tunneling photocurrent as a function of optical power, applied bias, and excitation photon energy. At high bias and high photon energy, the photocurrent grew linearly with optical power, while reducing either quantity caused the response to become progressively superlinear. The linear regime is consistent with direct photo-assisted tunneling: a large bias or high photon energy reduces the effective barrier sufficiently that primary photoexcited carriers tunnel before significant intralayer scattering occurs. The superlinear regime, by contrast, reflects thermionic emission: when the photon energy or bias is insufficient for direct tunneling, photoexcited carriers remain in the emitter long enough for carrier--carrier interactions to drive the electronic system toward a hot Fermi--Dirac distribution~\cite{song2011hot}, and only carriers in the high-energy tail of this distribution can overcome the barrier. By quantitatively analyzing the crossover between these two regimes, Ma et al.\ extracted the characteristic timescale of carrier--carrier scattering in graphene, demonstrating that the junction acts as an energy-selective gate whose threshold separates the ballistic tunneling and thermionic regimes of carrier relaxation.\\

The thermionic regime was examined in dedicated experiments by Massicotte et al.~\cite{Gr/WSe2/Gr_tunneling_thermionic} in graphene/WSe$_2$/graphene heterostructures, where the WSe$_2$ layer provides a well-defined barrier. Because graphene's carrier--carrier scattering is faster than carrier--phonon coupling~\cite{tielrooij2015generation}, absorbed photon energy rapidly thermalizes within the electronic system before being transferred to the lattice. Carriers in the high-energy tail of the resulting hot distribution are emitted over the barrier at the graphene/WSe$_2$ interface, generating photocurrent even for sub-bandgap photons~\cite{photothermionic_2}. In a related regime, Mylnikov et al.~\cite{Gr/BN/Gr_IR} identified a bolometric response in graphene/hBN/graphene junctions under mid-infrared illumination, where radiation-induced electron heating modifies the tunneling probability through localized defect states rather than driving over-barrier emission.\\

\subsection{Energy relaxation and bosonic modes}

Beyond carrier--carrier scattering, photoexcited carriers couple inelastically to bosonic excitations---phonons, plasmons, and polaritons---which act as discrete energy-exchange quanta during tunneling. When electron--boson coupling in the electrodes or barrier is significant, tunneling can proceed through boson absorption or emission, opening inelastic channels alongside elastic transport~\cite{tunneling_spectroscopy_book, inelastic_emitter_2019}. These processes appear as characteristic steps, thresholds, or resonances in tunneling spectra, providing spectroscopic fingerprints of specific vibrational or collective modes and their coupling strengths. In atomically sharp vdW junctions, where interfacial disorder is minimized and bosonic spectra are well defined, such inelastic signatures enable direct electrical readout of bosonic excitations that are otherwise typically resolved by optical spectroscopy~\cite{electricOptical_tunneling_review, ahmad2023progress, inelastic_emitter_2019}.\\

\begin{figure}
\centering
\includegraphics[width=6.5in]{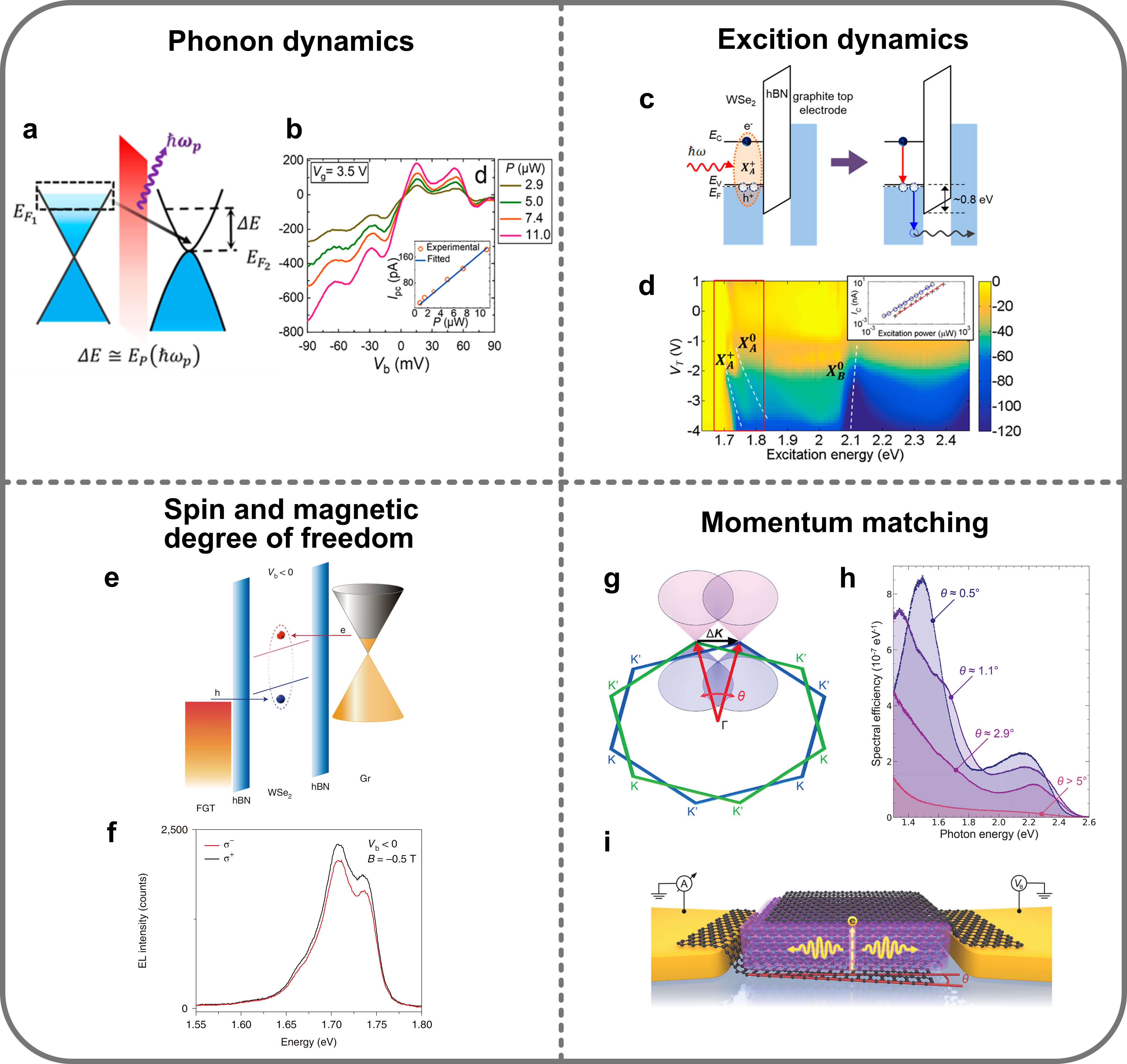}
\caption{\textbf{Photo-assisted tunneling as a probe of microscopic dynamics in vdW junctions.}
\textbf{(a,b)} \textit{Phonon dynamics:} inelastic tunneling assisted by phonons ($\hbar\omega_p$) produces
resonance-like photocurrent features versus bias, enabling extraction of mode energies and interfacial electron–phonon coupling. Adapted from Ref.~\cite{photon_phonon_assisted_tunneling}.
\textbf{(c,d)} \textit{Exciton dynamics:} optically generated excitons, including many-body or Auger-mediated pathways, feed energetic carriers into interlayer tunneling; excitation-energy and bias maps resolve excitonic resonances and decay pathways. Adapted from Ref.~\cite{WSe2/BN/Gr_Xiadong_Auger}.
\textbf{(e,f)} \textit{Spin and magnetic degrees of freedom:} magnetic order and optical polarization control spin-selective interlayer transfer, yielding helicity-resolved optical and electrical contrasts in tunneling-based readout. Adapted from Ref.~\cite{WSe2/BN/FGT_EL}.
\textbf{(g,h,i)} \textit{Momentum matching:} twist-controlled valley displacement $\Delta K$ sets momentum-matched tunneling pathways; spectral efficiency evolves strongly with twist angle, evidencing momentum-dependent photon-assisted tunneling. Adapted from Ref.~\cite{Gr_light_momentum_resonant}.
These panels show that photo-assisted tunneling can probe phonon, exciton, spin, and momentum-dependent dynamics at vdW interfaces.
}
\label{fig5}
\end{figure}

The phonon-assisted inelastic tunneling channel was directly probed in graphene-based vertical tunnel junctions by Xie et al.~\cite{photon_phonon_assisted_tunneling}, who measured the bias-dependent tunneling photocurrent at low temperature. They observed characteristic peak and valley features in the photocurrent that shifted weakly with gate voltage and gradually broadened with increasing temperature, as shown in Fig.~\ref{fig5}(b). By comparing the photocurrent spectrum with the second derivative of the independently measured dark tunneling current $d^2I_{dark}/dV^2$, they showed that the features are consistent with phonon-assisted inelastic electron tunneling, in which carriers exchange discrete energy quanta with lattice excitations during the tunneling event itself [Fig.~\ref{fig5}(a)], establishing the tunneling photocurrent as a sensitive electrical readout of electron--phonon coupling in the junction.\\

Mid-infrared excitation can additionally reveal high-energy optical phonon modes that are only weakly visible in dark transport. In the mid-infrared study by Mylnikov et al.~\cite{Gr/BN/Gr_IR} on graphene/hBN/graphene junctions, photocurrent maps showed gate-independent bias landmarks appearing as abrupt increases in photocurrent. The higher-bias feature occurs at an energy exceeding the incident photon energy while coinciding with graphene optical-phonon modes. Although the same resonances appear as weak local extrema in the dark $d^2I_\text{dark}/dV^2$ spectrum, they become clearly resolved in the photocurrent only under mid-infrared excitation. Taken with the low-energy phonon-assisted inelastic electron tunneling fingerprints above, these results establish Fig.~\ref{fig5}(b) as a bias-resolved spectroscopic readout of inelastic thresholds spanning both low- and high-energy lattice modes in the tunnel stack.\\

Beyond phonon-assisted processes, electrons can also lose energy to other bosonic modes during tunneling such as plasmonic modes and surface plasmon polaritons. For instance, Parzefall \textit{et al.} studied a vdW tunneling emitter based on a vertical metal/hBN/graphene junction and measured bias-driven light emission arising from inelastic tunneling events~\cite{inelastic_emitter_2019}. They showed that tunneling electrons can directly emit photons or excite surface plasmon polaritons, and that the accessible emission window is set directly by the applied bias through the tunneling energy budget ($\hbar\omega \leq eV$)~\cite{inelastic_emitter_2019}. They further demonstrated that coupling the same vdW tunnel junction to a nanocube cavity strongly reshapes and enhances the emitted spectrum, showing that the optical mode density and local nanophotonic environment are active control parameters for the tunneling-to-light conversion process, not merely secondary design details~\cite{ inelastic_emitter_2019}. Namgung \textit{et al.} reported a graphene tunnel-junction plasmonic emitter in a single-nanoparticle geometry~\cite{inelastic_blasmonic_emitter}, reinforcing the same physical picture that boson-assisted tunneling can be read out optically through the emitted spectrum and its bias dependence.\\

\subsection{Exciton dynamics}

In semiconductors---particularly in atomically thin transition metal dichalcogenides---optical excitation does not produce simple free carriers. The reduced dielectric screening and enhanced Coulomb interactions in two dimensions instead stabilize tightly bound electron--hole pairs known as excitons, with binding energies of several hundred meV~\cite{TMDs_Mak, chernikov2014exciton,rivera2015observation}. Because excitons are electrically neutral quasiparticles, they cannot tunnel as intact composite objects across a potential barrier in response to an applied bias~\cite{exciton_diss}. Their impact on vertical transport therefore arises indirectly: through dissociation into free electrons and holes, which are then individually subject to the tunneling potential, and through collective relaxation pathways that generate energetic free carriers. This makes the tunnel junction an electrical probe of exciton formation, binding, dissociation, and decay---complementary to optical probes such as photoluminescence and transient absorption spectroscopy~\cite{wang2018colloquium}.\\

A direct consequence of exciton formation is the strong and spectrally selective enhancement of optical absorption near excitonic resonances, which imprints corresponding features onto the tunneling photocurrent. This was demonstrated clearly in TMD-based vertical junctions, where the photocurrent spectrum reflects specific excitonic processes rather than the featureless absorption of a metallic emitter. For instance, in a graphene/hBN/WS$_2$/graphene heterostructure, Dong et al.~\cite{WS2/BN/Gr_bifunctional} illuminated the device with near-infrared photons well below the single-photon gap of WS$_2$ but resonant with its two-photon absorption window. By measuring the photocurrent as a function of excitation power and wavelength, they observed a quadratic power dependence and a photoresponse extending to $\sim$1145~nm, with resonant enhancement peaks at 1090~nm and 1145~nm corresponding to the 3p and 2p exciton states of WS$_2$. This indicates that two-photon-generated excitons dissociate at the WS$_2$/hBN interface and tunnel through the barrier before radiative recombination, converting a nonlinear optical process into an electrical signal. This kind of readout is difficult to obtain in lateral geometries.\\

At elevated photoexcitation densities, exciton--exciton interactions qualitatively modify the excitonic response and open additional tunneling pathways. In a MoS$_2$/hBN/graphene heterostructure, Linardy et al.~\cite{MoS2/BN/Gr_EEA} resonantly excited the ground exciton transition of monolayer MoS$_2$, and measured the photocurrent as a function of excitation power. An excited hole carries insufficient energy to overcome the $\sim$0.4~eV hole barrier of hBN, so the origin of the observed photocurrent is not immediately obvious. They found that the number of collected carriers per absorbed photon (internal quantum efficiency) drops at low power but saturates to a finite value at high excitation density, consistent with a process whose rate scales quadratically with exciton density. Combining these measurements with ab initio calculations, they identified exciton--exciton annihilation as the mechanism: collisions between exciton pairs generate hot holes with excess kinetic energies exceeding 1~eV, sufficient to overcome the hBN barrier. This demonstrates that vertical tunneling transport can directly probe nonradiative decay channels that are otherwise inferred only indirectly from photoluminescence quenching or transient absorption.\\

Auger recombination is another interaction-driven route to hot-carrier generation in tunnel junctions, depicted schematically in Fig.~\ref{fig5}(c). In this process, the recombination energy of one electron--hole pair is transferred to a third carrier rather than emitted as a photon. In WSe$_2$/hBN/graphene junctions, Chow et al.~\cite{WSe2/BN/Gr_Xiadong_Auger} measured the photocurrent spectrum as a function of excitation energy and applied bias. They observed sharp resonant enhancements of the photocurrent at the charged exciton (trion) transition energy of WSe$_2$, with a strongly asymmetric bias dependence, as shown in Fig.~\ref{fig5}(d). The resonant enhancement of the photocurrent at the trion transition energy of WSe$_2$ identifies Auger recombination as the dominant generation mechanism. The strongly asymmetric bias dependence separately reflects the asymmetric band offset between WSe$_2$ and hBN, which permits hot-hole emission while blocking hot electrons, making the device a direct electrical sensor of the Auger rate. In a complementary study, Sushko et al.~\cite{WSe2/BN/Gr_PhilipKim_Auger} examined hBN-encapsulated MoSe$_2$ and WSe$_2$ monolayers and measured substantial out-of-plane photocurrents under resonant optical excitation. By varying the hBN thickness and performing detailed ab initio modeling of the TMD/hBN band offsets, they confirmed that the asymmetric photoelectric response originates from Auger-generated hot holes escaping preferentially through the asymmetric barrier.\\

\subsection{Spin and magnetic degrees of freedom}

Optical excitation can manipulate spin and magnetic order through angular-momentum transfer from photons to carriers~\cite{2D_magnets_Burch,Gr/CrI3/Gr_vertical_photodiode,TMDs_Xiaodong}. In materials with broken inversion symmetry or valley-contrasting orbital moments, circularly polarized light can selectively excite carriers with specific spin or valley indices, generating polarized nonequilibrium populations~\cite{TMDs_Mak,MoS2_Mak,TMDs_DiXiao}. When such carriers participate in tunneling, the resulting current can become spin polarized, enabling optically controlled spin injection and detection across tunnel barriers~\cite{WSe2/BN/FGT_EL,InSe/BN/Gr_circular,Gr/CrI3/Gr_vertical_photodiode}.\\

One possible scenario is that circularly polarized light generates a spin-polarized carrier distribution in the emitter, while the tunneling barrier itself acts as a spin-sensitive filter that converts this polarization into a measurable current imbalance between the two light helicities. In InSe/hBN/graphene junctions, Pasquale et al.~\cite{InSe/BN/Gr_circular} illuminated few-layer InSe with circularly polarized light and measured the tunneling photocurrent as a function of the light helicity and out-of-plane magnetic field. Under an applied magnetic field, a helicity-dependent photocurrent emerged upon circular excitation. Because the out-of-plane $g$-factor of InSe is nearly zero, the relevant Zeeman splitting resides predominantly in the spin bands of the hBN barrier, making the barrier height itself different for different spins and enabling the spin polarization of the photocurrent to be read out directly from the tunneling current~\cite{InSe/BN/Gr_circular}.\\

The reverse scenario is also possible: a magnetic electrode injects a spin-polarized tunneling current, which is then detected optically through circularly polarized light emission. In transition metal dichalcogenides, spin--valley locking means that injecting spin-polarized carriers selectively populates one valley and therefore one circular polarization of luminescence. This was demonstrated in WSe$_2$/hBN/Fe$_3$GeTe$_2$ (FGT) heterostructures as shown in Fig.~\ref{fig5}(e), where the ferromagnetic FGT layer replaces the conventional non-magnetic electrode. Li et al.~\cite{WSe2/BN/FGT_EL} applied a bias to drive hole tunneling from FGT through hBN into WSe$_2$ and measured the helicity of the resulting electroluminescence as a function of applied magnetic field. They observed that the circular polarization of the emitted light traced the magnetic hysteresis loop of the FGT contact and reversed sign upon magnetic field reversal, as illustrated in Fig.~\ref{fig5}(f). This indicates that the spin orientation of holes injected from FGT is transferred directly into valley polarization in WSe$_2$ via spin--valley locking, providing an all-electrical route to valleytronic control without optical pumping.\\

Whether spin selectivity originates from optical helicity in the emitter or from spin-polarized injection via a ferromagnetic contact, the tunneling current captures the conversion between spin-dependent transport and optical polarization, and vice versa. Because spin, valley, and excitonic degrees of freedom are coupled in these materials~\cite{TMDs_Xiaodong, TMDs_DiXiao, 2D_magnets_Burch}, the tunneling readout reflects their joint nonequilibrium dynamics.

\subsection{Ferroelectric tunnel junctions}

Ferroelectric polarization provides an additional route to control tunneling transport by directly reshaping the electrostatic potential across the junction. Unlike optical excitation, which primarily modifies carrier populations in the electrodes, ferroelectric polarization acts from within the barrier itself, altering the local potential profile, interfacial band alignment, and therefore the tunneling transmission probability. This mechanism underlies ferroelectric tunnel junctions, in which an ultrathin ferroelectric layer serves as the tunneling barrier between two electrodes.\\

In a ferroelectric tunnel junction, spontaneous polarization generates bound surface charges at the interfaces and produces an internal electric field that modifies the effective barrier height and width experienced by tunneling carriers. Reversal of the polarization direction therefore leads to distinct conductance states, giving rise to tunneling electroresistance (TER). Because the polarization state can be switched in a non-volatile manner by an external electric field, the tunneling conductance itself becomes a memory variable encoded in the ferroelectric configuration. A representative example is the vdW ferroelectric tunnel junction, such as CuInP$_2$S$_6$/graphene heterostructure~\cite{wu2020high,wang2022extraordinary}, where the layered ferroelectric CuInP$_2$S$_6$ serves as the tunneling barrier. Polarization reversal in CuInP$_2$S$_6$ strongly modulates the graphene Fermi level due to its low density of states, leading to a large change in the effective barrier height and a giant tunneling electroresistance exceeding $10^7$. This work highlights how ferroelectric polarization in vdW junctions can control tunneling transport not only through the barrier itself but also via electrostatic reconstruction of adjacent electrodes.\\

When combined with optical excitation, polarization-controlled tunnel junctions enable additional functionalities by converting the polarization-induced internal field into a driving force for photocarrier transport. A representative example is the graphene/3R-MoS$_2$/graphene vertical junction~\cite{yang2022spontaneous}. In this system, rhombohedral stacking in few-layer MoS$_2$ produces an intrinsic out-of-plane polarization that generates a depolarization field across the junction. Under illumination, this built-in field separates photoexcited carriers and drives them toward opposite graphene electrodes through tunneling, producing a zero-bias photovoltaic current. Owing to the atomically thin geometry and weak screening in graphene, the device achieves efficient photocarrier extraction with an external quantum efficiency reaching $\sim$16\%, demonstrating how intrinsic polarization in vdW materials can directly control photo-assisted tunneling transport.\\

Polarization can further couple with magnetic order to form multiferroic tunnel junctions, in which ferroelectric polarization and electrode magnetization jointly control tunneling transport. In these devices, polarization reversal modifies the barrier profile, whereas the relative magnetization alignment governs spin-dependent transmission, giving rise to the coexistence of tunneling electroresistance and tunneling magnetoresistance. This concept was first proposed theoretically in vdW heterostructures combining ferromagnetic electrodes with a ferroelectric barrier~\cite{su2020van}. More recently, it was realized experimentally in all-vdW junctions such as Fe$_3$GeTe$_2$/CuInP$_2$S$_6$/Fe$_3$GeTe$_2$, where four distinct non-volatile resistance states were demonstrated~\cite{xie2026tailorable}. These results establish polarization-coupled magnetic tunneling as a promising route toward multistate memory and spintronic functionalities.\\

\subsection{Momentum matching and twist-controlled tunnel junctions}

In clean vdW tunnel junctions, in-plane momentum conservation is an additional selection rule beyond energy matching: an electron can only tunnel if a state of the same in-plane wavevector is available on the other side of the barrier. This constraint is most consequential when both electrodes share the same crystal symmetry and nearly identical lattice constants, so that their Brillouin zones are nearly congruent in reciprocal space. In that case---exemplified by graphene/hBN/graphene stacks---a twist by $\theta$ displaces the K points of one electrode relative to the other by $|\Delta K| = 2K\sin(\theta/2)$, which vanishes at alignment ($\theta = 0$) and provides a single, continuously tunable momentum mismatch~\cite{resonant_tunneling_2d, Gr_resonant2}, as depicted schematically in Fig.~\ref{fig5}(g).\\

In many practical junctions---especially when the emitter and collector are composed of different materials with incommensurate lattice periodicities, or when identical materials are operated far from the aligned limit---the dominant transfer channels are momentum-mismatched ones mediated by disorder or phonons~\cite{chandni2016signatures, chandni2015evidence}. This is why the junctions discussed in preceding sections were treated within an energy-selective tunneling picture without a sharp angular selection rule. However, even in such junctions the relative crystallographic orientation is not irrelevant: although exact global momentum matching is generally absent in the operating configuration, the twist angle can still strongly modify interlayer coupling, the overlap of available momentum-space states, and partial commensuration conditions, thereby substantially changing carrier transfer efficiency~\cite{chen2022momentum, QTM, QTM_phonon, guo2025anisotropic, srivastava2021resonant}.\\

The impact of this selection rule on photodetection is direct. In aligned graphene/hBN/graphene vertical tunnel junctions, Xie et al.~\cite{ xie2024resonant} systematically tuned the crystallographic orientation of the two graphene electrodes and measured the tunneling photocurrent as a function of bias and gate voltage. They observed a two-orders-of-magnitude enhancement in photocurrent in the aligned device compared to the misaligned case. The aligned device also exhibited negative differential conductance: at a specific bias the Dirac cones of the two graphene layers reach optimal momentum alignment, and increasing the bias beyond this point shifts them out of alignment, reducing the current. Both observations support momentum-conserving resonant tunneling as the dominant photoresponse mechanism and account for the measured external photoresponsivity of 0.7~A/W across a wide spectral and temperature range.\\

The same momentum constraints are directly visible in optical tunneling spectra. In graphene/hBN/graphene light-emitting junctions, Kuzmina et al.~\cite{Gr_light_momentum_resonant} fabricated devices with systematically varied twist angles and measured the electroluminescence spectrum as a function of $\theta$ and applied bias. Near alignment ($\theta \lesssim 3^\circ$), they observed a sharp resonant emission peak at $\hbar\omega \approx \Phi$, where $\Phi$ is the bias-controlled energy offset between the Dirac cones, with a high-energy cutoff at $\hbar\omega_\mathrm{cut} \approx eV$, where $e$ is the elementary charge and $V$ is the applied bias voltage. As $\theta$ was increased toward $\sim$1.1$^\circ$, $\sim$2.9$^\circ$, and beyond, this resonant feature progressively broadened and weakened, eventually vanishing for strong misalignment, as illustrated in Fig.~\ref{fig5}(h). This indicates that the resonant peak originates from momentum-conserving photon-assisted cone-to-cone transitions whose phase space shrinks with increasing valley displacement, while the residual broadband background reflects momentum-relaxing inelastic channels that dominate once $\Delta K$ is large.

\section{Emergent applications enabled by light-controlled tunnel junctions}
\label{sec:applications}

Having established the microscopic mechanisms of photo-assisted tunneling in the preceding sections, we now turn to the functional capabilities that emerge from this barrier-defined paradigm. Unlike conventional photodiodes, where spectral response and transport are constrained by the absorber bandgap and drift--diffusion across micrometer-scale depletion regions, tunneling-based architectures rely on barrier transmission as the primary energy filter and quantum-mechanical penetration across atomically thin barriers for carrier extraction. This barrier-defined operation introduces spectral, electrostatic, temporal, and quantum-mechanical degrees of control that are inaccessible in bulk semiconductor optoelectronics. By leveraging energy-selective tunneling, nonlinear transport, and electrostatic tunability, vdW tunnel junctions could overcome long-standing bottlenecks in photodetection, optical communication, and information processing.\\

In the following subsections, we highlight representative advances in (i) advanced photodetection and photosensing, (ii) novel tunnel junction-based light sources and emission control, and (iii) memory, in-sensor computing, and neuromorphic applications enabled by light-controlled tunnel junctions.

\subsection{Advanced photodetection and photosensing}
\label{subsec:spectral_app}

Light-controlled tunnel junctions offer a fundamentally different paradigm for photodetection and optical sensing. In conventional photonic and electronic detectors, key performance metrics including spectral response, noise floor, gain, and speed are intrinsically constrained by the absorber bandgap and by drift--diffusion transport across micrometer-scale depletion regions~\cite{nano_photodetector, konstantatos2010nanostructured}. In tunneling-based architectures, the barrier acts as the primary energy filter, and because its height is electrostatically programmable in situ, the spectral response, gain, and speed can be tuned more independently rather than being fixed by the choice of absorber material. This barrier-defined operation enables ultrafast carrier extraction, strong suppression of thermally activated dark current, and independent electrostatic control of spectral selectivity, introducing spectral, temporal, and quantum-mechanical degrees of control that are inaccessible in bulk semiconductor photodetectors~\cite{electricOptical_tunneling_review, tunneling_in_vdw_book}. As a result, tunneling-based devices overcome long-standing trade-offs between sensitivity, speed, and noise, positioning light-controlled tunnel junctions as a versatile and electrically reconfigurable platform for next-generation photosensing across a wide spectral range~\cite{ahmad2023progress}.

\subsubsection{Barrier-defined spectral selectivity}

\begin{figure}
\centering
\includegraphics[width=6.5in]{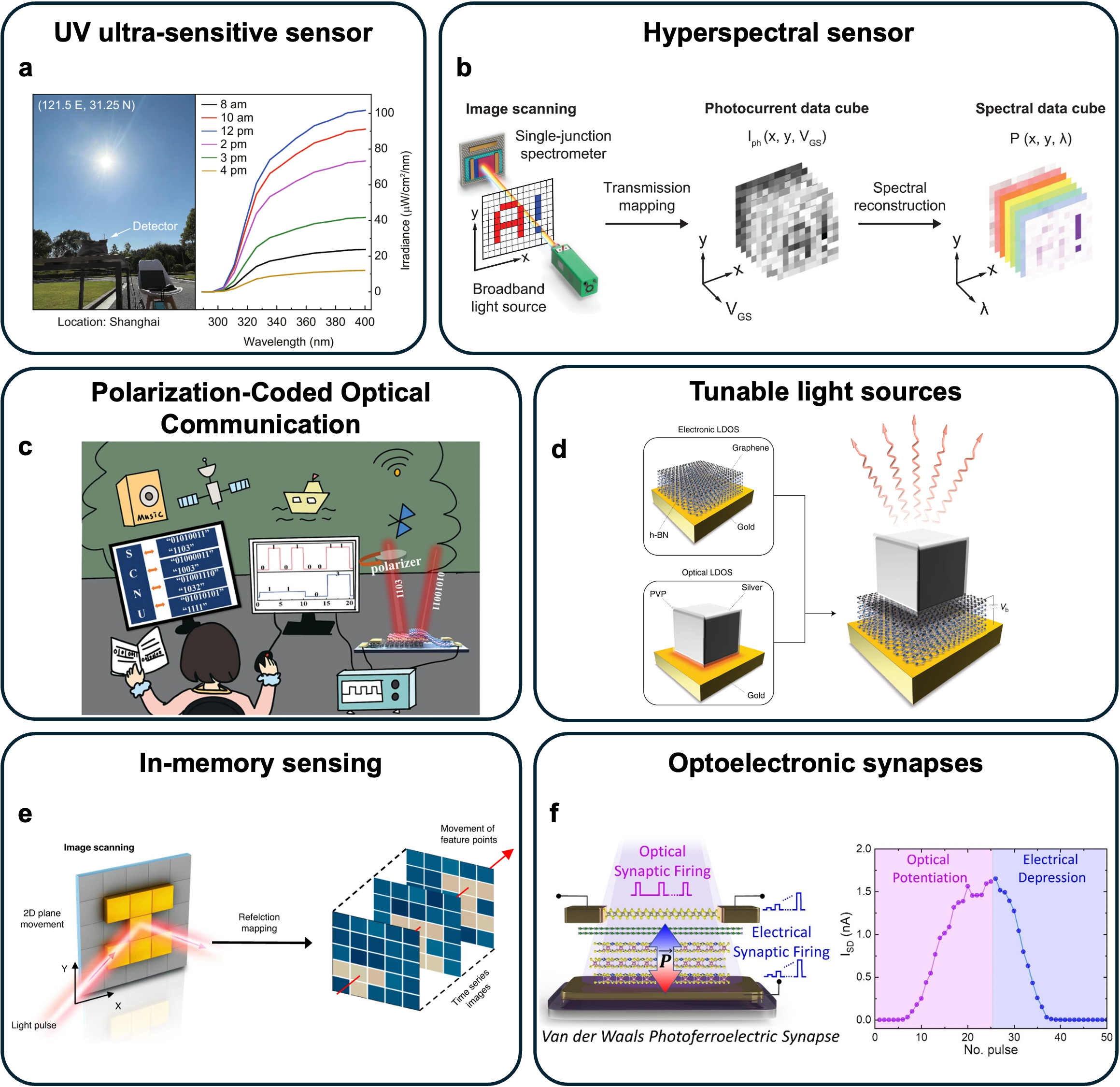}
\caption{
\textbf{Emergent applications enabled by light-controlled tunneling.}
\textbf{(a)} Solar-blind UV sensing via barrier-defined energy filtering, with preferential extraction of high-energy carriers and suppression of visible-background response. Adapted from Ref.~\cite{MoS2/BN/Gr_UV}.
\textbf{(b)} Single-junction computational spectroscopy and hyperspectral sensing: electrically reconfigurable photocurrent fingerprints $I_{\mathrm{ph}}(x,y,V_{\mathrm{GS}})$, where $x$ and $y$ denote spatial position and $V_{\mathrm{GS}}$ is the gate voltage, are reconstructed into spectral data cubes $P(x,y,\lambda)$, where $\lambda$ is the wavelength and $P$ is the reconstructed spectral intensity, without dispersive optics. Adapted from Ref.~\cite{MoS2/WSe2_mini_spec}.
\textbf{(c)} Polarization-coded optical communication enabled by polarization-selective tunneling photoresponse in anisotropic vdW heterostructures. Adapted from Ref.~\cite{ReSe2/SnSe2}.
\textbf{(d)} Tunable tunneling-based light sources, illustrating electrically programmable emission in the reciprocal tunnel-junction operating mode. Adapted from Ref.~\cite{inelastic_emitter_2019}.
\textbf{(e)} In-memory sensing and motion-aware front-end processing, where sensing, state retention, and early computation are co-localized in tunneling hardware. Adapted from Ref.~\cite{BP/InSe_B2B}.
\textbf{(f)} Optoelectronic synaptic behavior, including potentiation- and depression-like multilevel updates, for neuromorphic processing. Adapted from Ref.~\cite{optoelectronic_synapse}.
}
\label{fig6}
\end{figure}

A direct consequence of barrier-defined spectral filtering is the ability to realize deep-ultraviolet (DUV) photodetection in a device whose absorber, graphene, has no bandgap at all. DUV detection conventionally requires bulk wide-bandgap semiconductors such as AlGaN or Ga$_2$O$_3$, which are difficult to grow and p-type dope~\cite{xu2019gallium}. In MoS$_2$/hBN/graphene tunnel junctions, the hBN barrier passes only high-energy photoexcited carriers while blocking visible and infrared electrons from the gapless graphene layer, yielding a responsivity of $4.4 \times 10^6$~A/W, a specific detectivity of $1.4 \times 10^{17}$~Jones at 250~nm, and a DUV-to-visible rejection ratio exceeding $10^6$~\cite{MoS2/BN/Gr_UV}, as shown in Fig.~\ref{fig6}(a). The barrier-defined cutoff is material-agnostic: replacing hBN with Al$_2$O$_3$ or the graphene absorber with PdSe$_2$ or metal films reproduces the same solar-blind selectivity without redesigning the device geometry~\cite{ MoS2/BN/Gr_UV}. For applications such as flame sensing, missile plume detection, and UV-C sterilization verification, where rejection of visible and infrared background is essential, this built-in spectral filtering eliminates the need for external bandpass optics entirely.\\

Crucially, the spectral response in tunnel junctions can be dynamically reconfigured by the applied electrostatic bias and gate rather than being fixed by material properties, hence the same junction that performs barrier-defined photodetection can be reprogrammed in situ into a computational spectrometer. In MoS$_2$/WSe$_2$ heterojunctions~\cite{MoS2/WSe2_mini_spec}, sweeping the gate voltage modifies the interlayer band alignment, producing a series of voltage-dependent spectral fingerprints from which the incident spectrum can be reconstructed [Fig.~\ref{fig6}(b)]. This single-pixel approach achieves a wavelength accuracy of 0.36~nm, a spectral resolution of $\sim$3~nm, and an operating bandwidth spanning 405--845~nm, with resolution comparable to benchtop grating instruments but without dispersive optics, filter arrays, or moving parts~\cite{yang2021miniaturization}. Electrically programmable spectroscopy at the single-pixel level is directly relevant to on-chip chemical analysis, portable environmental monitoring, and implantable diagnostics~\cite{hyperspectral_imaging_review, geometric_optical_sensing_review}.\\

The same barrier-controlled concept extends naturally to longer wavelengths, where photon energies fall below the nominal barrier height yet still drive a measurable tunneling photocurrent through hot-carrier physics. In graphene/WSe$_2$/graphene heterostructures, absorbed photon energy thermalizes into the graphene electron bath within $\sim$50~fs before reaching the lattice; the resulting hot-carrier tail is emitted over the WSe$_2$ Schottky barrier via the photothermionic effect, enabling spectrally flat, electrically tunable detection out to 1500~nm --- well below the WSe$_2$ bandgap of $\sim$1.3~eV --- with a gate-tunable responsivity reaching 0.12~mA/W and a signal recovery time on the order of picoseconds~\cite{Gr/WSe2/Gr_tunneling_thermionic}. The Schottky barrier height, and therefore the detection efficiency, is continuously tunable by gate voltage, providing an electrical handle on the spectral response that has no counterpart in fixed-bandgap photodiodes. In graphene/hBN/graphene junctions, mid-infrared radiation induces electron heating that is transduced directly into a tunneling current, extending detection to wavelengths far below the hBN bandgap with a photocurrent proportional to the second derivative of the tunnel current --- a signature of resonant tunneling through hBN defect states~\cite{Gr/BN/Gr_IR}. At the mid-wave infrared (MWIR, 3--5~$\mu$m), MoS$_2$/BP heterojunction devices achieve room-temperature blackbody responsivity of 1.68~A/W in a 2D architecture~\cite{MoS2/BP}, a performance regime that conventional MWIR photodetectors reach only with cryogenic cooling of InSb or HgCdTe focal-plane arrays~\cite{rogalski2002infrared}. Collectively, these platforms extend the tunneling detection window continuously from 250~nm to beyond 5~$\mu$m within a single device family.

\subsubsection{Overcoming fundamental trade-offs in gain, speed, and noise}

A well-known constraint in photodetector design is the gain--bandwidth trade-off: high gain requires long carrier lifetimes, which directly degrades response speed, and the two metrics have historically moved in opposite directions~\cite{nano_photodetector}. Tunneling architectures circumvent this because photoinduced charges modulate the barrier height itself, accelerating carrier extraction independently of lifetime. This principle is demonstrated across a range of material combinations. In WSe$_2$/Ta$_2$NiSe$_5$ heterostructures, photogating-assisted tunneling delivers responsivities around $10^4$~A/W at microsecond response times, a combination that is difficult to obtain in conventional phototransistors or standard photodiodes~\cite{WSe2/Ta2NiSe5}. The same logic extends to hybrid 2D/bulk architectures: a Si/MoS$_2$/graphene tunneling photoconductor achieves $\sim$$3 \times 10^4$~A/W with a rise time of only 17~ns, three orders of magnitude faster than the equivalent device without the MoS$_2$ tunneling interlayer, because the atomically thin MoS$_2$ film simultaneously acts as a tunneling layer and passivates surface states that would otherwise trap carriers and slow the response~\cite{Si/MoS2/Gr_fast_large}. Engineering the tunneling interlayer for impact ionization rather than simple passivation offers a further handle: inserting an AlN barrier into a graphene/silicon junction promotes photocurrent multiplication through impact ionization with a photogain of $\sim$5.8 at 850~nm, while simultaneously suppressing the dark current by an order of magnitude relative to the bare graphene/silicon device, reaching 3.96~A/W~\cite{Si/AlN/Gr_photodetector}. Taken to the system level, graphene/hBN/graphene tunneling photodiodes monolithically integrated on Si$_3$N$_4$ waveguides demonstrate GHz bandwidth~\cite{Gr/BN/Gr_Si3N4_waveguides}, confirming that this gain--bandwidth performance survives the constraints of photonic chip integration and is not limited to isolated flake devices.\\

A persistent challenge in broadband photodetection is minimizing dark current without compromising responsivity. Tunneling architectures address this by employing large-bandgap physical barriers or engineered band offsets to block thermally activated carriers while remaining transparent to photoexcited carriers with sufficient energy~\cite{electricOptical_tunneling_review, tunneling_in_vdw_book}. In graphene/hBN/WS$_2$/graphene heterostructures, the inserted hBN tunneling barrier directly suppresses dark current, yielding a visible-range on/off ratio of $1.02 \times 10^3$; the same device extends photodetection into the near-infrared to 1175~nm through two-photon absorption enhanced by the WS$_2$ exciton resonance~\cite{WS2/BN/Gr_bifunctional}. Junctions with broken-gap band alignment enforce dark current suppression through band-offset engineering rather than a physical barrier: PtS$_2$/WSe$_2$ heterodiodes reach a rectification ratio approaching $10^8$ and a photocurrent on/off ratio above $10^5$~\cite{PtS2/WSe2}; WSe$_2$/SnS$_2$ tunneling photodiodes achieve a photocurrent on/off ratio of $\sim$$10^6$~\cite{WSe2/SnS2}; Bi$_2$O$_2$Se/SnSe$_2$ junctions reach an on/off ratio of $8.42 \times 10^4$, sustained under intense illumination by a hole potential well that suppresses carrier recombination while keeping leakage current low~\cite{ Bi2O2Se/SnSe2}; and Te/SnS$_2$ junctions exhibit a current on/off ratio exceeding $10^4$ with a reverse rectification ratio of $\sim$5000~\cite{Te/SnS2}. These results establish that barrier-controlled carrier selectivity, whether through a physical insulating layer or an engineered band offset, is a reliable route to achieving high photocurrent on/off ratios that are difficult to reach in conventional p-n photodiodes limited by thermally activated leakage.\\

The same band alignment that suppresses dark current simultaneously enables zero-bias, self-powered operation, an important practical advantage for deployments where an external power supply is unavailable. In junctions with staggered or broken-gap band alignment, the built-in interfacial potential drives efficient photocarrier separation without any applied bias, and the suppression of dark current by the barrier ensures that the resulting signal-to-noise ratio remains high. Te/SnS$_2$ tunneling junctions deliver 2.21~A/W and an external quantum efficiency of 678\% at zero bias~\cite{Te/SnS2}; PtS$_2$/WSe$_2$ broken-gap heterodiodes combine a rectification ratio approaching $10^8$ with passive photodetection at room temperature~\cite{PtS2/WSe2}. These results establish self-powered tunneling photodetectors as practical candidates for wearable health monitors, implantable biosensors, and distributed internet of things nodes where battery constraints are severe.

\subsubsection{Polarimetric and multidimensional sensing}

Tunneling junctions built from structurally anisotropic 2D materials deliver intrinsic polarimetric response without any external polarizing optics. The key figure of merit is the photocurrent anisotropy ratio, the contrast between photocurrents collected along orthogonal polarization axes, which in most single-material 2D polarization sensors is constrained to $\sim$2 by the modest structural anisotropy of available crystals~\cite{ma2024low}. In low-symmetry crystals, this direction-dependent photocurrent can arise from anisotropic absorption in the emitter layer, anisotropic collection in the collector layer, or both~\cite{WS2/Te, liu2025high, ReSe2/SnSe2}. Tunneling junctions preserve this contrast by suppressing interface trap states that would otherwise scatter carriers isotropically and wash out the polarization selectivity~\cite{WS2/Te, ReSe2/SnSe2}. Several material combinations illustrate this across an increasing range of anisotropy ratios: WS$_2$/Te heterostructures, where the anisotropic Te absorber drives the contrast, reach a ratio of 2.5~\cite{WS2/Te}; Bi$_2$O$_2$Se/Ta$_2$NiSe$_5$ junctions, exploiting the monoclinic chain structure of Ta$_2$NiSe$_5$, reach 2.12~\cite{yang2025bi2o2se}; MoS$_2$/PdSe$_2$ heterostructures with straddling band alignment, where the bandgap of one layer is entirely contained within that of the other, achieve an anisotropy ratio of 1.2 even with a structurally symmetric absorber, suggesting that the anisotropic PdSe$_2$ collector alone is sufficient to imprint polarimetric sensitivity across a broadband 350--1350~nm window~\cite{liu2025high}; and ReSe$_2$/SnSe$_2$ broken-gap junctions, where both layers are anisotropic, push the ratio to 12.26--13.27, approximately four times that of pristine ReSe$_2$ alone ($\sim$3.1), while operating at zero applied bias with a light on/off ratio of $10^5$~\cite{ ReSe2/SnSe2}, and have been demonstrated in polarization-coded optical communication at quaternary logic states without any power supply~\cite{ReSe2/SnSe2} [Fig.~\ref{fig6}(c)]. \\

Beyond linear polarization, tunneling junctions also respond to the circular polarization state of incident light: in InSe/hBN heterostructures, circularly polarized excitation generates an asymmetric tunneling photocurrent whose sign encodes light helicity, electrically detecting circular polarization without any waveplate or polarizing element~\cite{InSe/BN/Gr_circular}. Polarization sensitivity also emerges in heterostructures not primarily designed for it: WSe$_2$/Ta$_2$NiSe$_5$ tunneling devices show a clear dependence of photocurrent on the polarization state of incident light, adding a polarimetric dimension to their high-responsivity operation~\cite{WSe2/Ta2NiSe5}. In polarimetric imaging, remote sensing through scattering media, and polarization-resolved biomedical diagnostics, where conventional polarization-selective optical stacks multiply component count and alignment requirements~\cite{polarimetric_imaging_review, polarisation_optics_for_bio_review}, the monolithic polarimetric response of these junctions offers a route to compact, chip-integrated multidimensional sensing that extracts spectral, intensity, and polarization information simultaneously within a single device~\cite{wang2024multidimensional}.\\

The capabilities surveyed in this section reflect a qualitative shift in what a single device platform can do. Barrier-defined spectral control spanning the deep ultraviolet to the mid-infrared, simultaneous high gain and fast response, passive zero-bias operation, and monolithic polarimetric sensitivity are sensing modalities that are structurally inaccessible to conventional drift-diffusion photodetectors. Light-controlled tunneling thus positions vdW heterostructures as a foundational architecture for next-generation optical sensing and information extraction~\cite{electricOptical_tunneling_review, wang2024multidimensional, geometric_optical_sensing_review}.

\subsection{Novel tunneling-based light sources and emission control}
\label{subsec:lightsource_app}

The intrinsic reversibility of quantum tunneling implies that the same vdW tunnel junctions used for photodetection and sensing can also operate in the opposite mode, as electrically driven light sources. When biased to inject carriers into an emissive layer, tunneling junctions function as light-emitting diodes (LEDs) in which radiative recombination is governed by energy- and symmetry-selective carrier injection through the barrier. Unlike conventional LEDs, where emission characteristics are largely fixed by the band structure of the active material, tunneling-based emitters provide bias-dependent spectral selectivity, access to nonequilibrium radiative pathways, and direct control over spin, valley, and polarization degrees of freedom.

\subsubsection{Excitonic electroluminescence and spin-valley polarization control}

A demonstration of tunneling-injected electroluminescence is the WSe$_2$/hBN/graphene light-emitting tunneling transistor, in which carriers are injected through the hBN barrier into the WSe$_2$ active layer, producing room-temperature electroluminescence with brightness substantially exceeding that of conventional WSe$_2$ photodiodes at equivalent injection current~\cite{WSe2/hBN/Gr_light_emitter}. The enhancement arises because tunneling injection deposits carriers directly into the radiative excitonic states of the TMD without passing through a bulk depletion region, reducing non-radiative losses at the interface. The detection--emission duality enabled by the same barrier is directly illustrated in graphene/hBN/WS$_2$/graphene heterostructures, where the device functions as a photodetector at low bias and transitions to electroluminescence at high bias, with the gate voltage tuning the proportion of neutral and charged excitons in the emissive WS$_2$ layer~\cite{WS2/BN/Gr_bifunctional}.\\

Spin and valley degrees of freedom in the emissive TMD layer are 
also electrically addressable in this tunneling architecture. In WSe$_2$/hBN/Fe$_3$GeTe$_2$ heterostructures, the ferromagnetic Fe$_3$GeTe$_2$ contact injects spin-polarized holes through the hBN tunnel barrier into WSe$_2$, selectively populating specific spin--valley states and producing circularly polarized electroluminescence~\cite{WSe2/BN/FGT_EL}. The degree and handedness of the circular polarization follow the magnetic hysteresis loop of the Fe$_3$GeTe$_2$ layer and can be reversed by an external magnetic field or tuned electrically, providing dynamic and reversible control over the polarization state of the emitted light that is not available in conventional electroluminescent devices. Such electrically programmable spin-LEDs offer a compact platform for polarization-encoded light sources with applications ranging from optical interconnects to valleytronic optoelectronics~\cite{TMDs_Xiaodong, TMDs_DiXiao}.

\subsubsection{Inelastic tunneling as a direct photon emission mechanism}

Beyond excitonic electroluminescence, photons can be generated directly by the tunneling event itself through inelastic electron tunneling, in which an electron crossing the barrier loses energy to a photon rather than to a phonon. In graphene/hBN/graphene tunnel junctions, applying a bias drives inelastic tunneling transitions that emit broadband visible and near-infrared radiation, with the emission spectrum set by the applied voltage window and continuously tunable by bias~\cite{inelastic_emitter_2019} [Fig.~\ref{fig6}(d)]. This voltage-defined spectral control is independent of any material bandgap, making the emission energy a direct electrical degree of freedom. The same inelastic tunneling mechanism can be coupled to plasmonic resonances: in graphene tunnel junctions integrated with metallic nanoparticles, inelastic tunneling excites localized plasmons that radiate as a nanoscale light source with a footprint below the diffraction limit~\cite{inelastic_blasmonic_emitter}. Together, these results establish that vdW tunnel junctions can function as bias-programmable, nanoscale light sources whose emission does not require an excitonic or bandgap transition.

\subsubsection{Momentum- and bias-controlled electroluminescence}

The graphene/hBN/graphene platform, most commonly associated with tunneling transport and photodetection, can also function as a light source. When momentum conservation in the tunneling transition is enforced by the crystallographic alignment between graphene electrodes, the emission becomes spectrally narrow and twist-angle-tunable. In graphene/hBN/graphene junctions, twist-controlled resonant tunneling produces a spectral emission peak in the near-infrared that is attributed to photon-assisted momentum-conserving electron tunneling; rotating the relative crystal orientation shifts the peak by over 0.2~eV, providing a structural handle on emission wavelength~\cite{Gr_light_momentum_resonant}. Nearly aligned devices show pronounced resonant features in both optical and electrical characteristics that vanish rapidly for twist angles above $\sim$3°, directly linking the emission spectrum to the interlayer commensurability. At the opposite spectral extreme, the same graphene/hBN/graphene platform produces deep-ultraviolet electroluminescence when driven at high bias: tunneling carriers injected into the hBN band edges activate radiative recombination channels at $\sim$215~nm, well beyond the reach of conventional nitride-based DUV emitters that require complex epitaxial growth and p-type doping~\cite{Gr-hBN-Gr-UV}.\\

Light-controlled tunneling thus enables a class of optoelectronic light sources in which spectral output, polarization state, and temporal response are dictated by barrier design and electrostatic bias rather than by intrinsic material constraints. The tunnel barrier acts as an active filter that shapes the energy, spin, and momentum of injected carriers, determining when, where, and how radiative recombination occurs and allowing emission properties to be programmed in situ within a single compact device.

\subsection{Memory, in-sensor computing, and neuromorphic applications}

Light-controlled tunneling extends naturally beyond instantaneous photodetection: when photoexcited carriers are trapped at a barrier interface or in a floating gate, the same junction that detects light also retains information about it, acquiring persistent conductance states that encode optical exposure history~\cite{ju2014photoinduced, liao2022bioinspired}. Because sensing, storage, and computation can occur in the same nanoscale device, light-controlled tunneling devices can reduce data movement penalties of sensor-to-memory transfer that usually dominate conventional machine-vision systems.\\

\textbf{(1) Optoelectronic memory enabled by photogating.} The simplest realization is floating-gate optoelectronic memory, where photoexcited hot carriers tunnel through a barrier and are retained in a charge-trapping layer, programming the device into one of several well-separated conductance levels that persist in the dark. In PtS$_2$/hBN/graphene heterostructures, asymmetric hot-carrier tunneling through the hBN barrier stores charge in the graphene floating gate, producing multiple distinguishable memory states that record optical exposure non-destructively~\cite{PtS2/BN/Gr}. Multibit storage is further enabled by combining optical sensitivity with engineered charge-trapping sites: MoS$_2$-based floating-gate devices incorporating Au nanoparticle charge-trap layers achieve ultrahigh photoresponsivity alongside multiple resolved conductance levels~\cite{MoS2/cPVP/AuNPs_Multibit}, demonstrating that the number of accessible memory states is controlled by trap density engineering rather than fixed by the material bandgap. Memory functionality can also be engineered on faster timescales: in band-to-band tunneling devices based on BP/InSe heterostructures, the nearly infinitesimal tunnel barrier facilitates sub-microsecond photomemory retention --- fast enough to detect moving targets without motion blur while suppressing static backgrounds~\cite{BP/InSe_B2B} [Fig.~\ref{fig6}(e)].\\

A qualitatively distinct memory mechanism is offered by ferroelectric tunnel junctions, where the stored information resides in the polarization state of the tunnel barrier itself rather than in trapped charge. Interfacial two-dimensional ferroelectricity arising from sliding in twisted TMD bilayers produces switchable spontaneous polarization that modulates the tunneling transmission, enabling ambipolar resistive switching and well-defined remanent states~\cite{gao2024tunnel}. When the ferroelectric barrier is embedded in a photosensitive vdW heterostructure, optical excitation gains direct control over the polarization state, combining photodetection with nonvolatile ferroelectric memory in a single stack. In ReS$_2$/hBN/CuInP$_2$S$_6$ junctions, the ferroelectric polarization of the CuInP$_2$S$_6$ layer can be written electrically or optically through band-to-band photocarrier generation in ReS$_2$ and photovoltaic response in the ferroelectric, achieving an on/off ratio exceeding $10^7$, a hysteresis memory window up to 7~V wide, and retention exceeding $10^3$~s, while supporting three distinct synaptic operating modes: electrically stimulated, optically stimulated, and optically assisted~\cite{ optoelectronic_synapse} [Fig.~\ref{fig6}(f)]. This positions ferroelectric vdW tunnel junctions as a functionally richer memory platform than charge-trapping devices, with optical writing and electrical reading as independent operations.\\

\textbf{(2) In-sensor computing and spectral processing.} Beyond memory, light-controlled tunneling enables in-sensor computing, where optical signals are processed directly at the detector level rather than being transmitted to a separate processor. Gate-tunable PdSe$_2$/MoTe$_2$ junctions can be electrostatically reconfigured between staggered band alignment and broken-gap alignment. This reconfigurability continuously tunes both the sign and magnitude of the photoresponse, performing broadband analog convolution entirely within the detector front end, with recognition accuracy demonstrated on multi-band infrared image datasets~\cite{PdSe2/MoTe2_processing}. The linear conductance programmability demonstrated in MoS$_2$-based floating-gate architectures~\cite{MoS2/BP, tang2021reliable}, where device weights are updated proportionally to optical pulse power at an energy cost of 1.8~fJ per programming event, is a direct analogue of the weight-update rule required for hardware neural network training. A complementary route to in-sensor computing operates in the spectral domain: electrically reconfigurable tunnel junctions such as MoS$_2$/WSe$_2$ can sweep their tunneling transmission window across a broad spectral range, generating a sequence of voltage-programmed spectral projections that serve as the analog spectral encoding layer of an in-sensor convolutional processor~\cite{MoS2/WSe2_mini_spec}. This is conceptually analogous to integrating spectral filter arrays directly on image sensors to perform analog convolution in the wavelength domain before any digital processing, an approach recently shown to reduce downstream data throughput by over 96\% while achieving $>$96\% accuracy on real-world classification tasks without GPU-based inference~\cite{cui2025spectral}. In the tunneling implementation, the spectral weights are not fixed by fabrication but are electrically programmable in situ, offering a reconfigurability that static filter arrays cannot provide.\\

\textbf{(3) Neuromorphic and synaptic architectures.} Nonvolatile optical weight storage, analog convolution, and spectral front-end processing are each individually compelling, but they are also constituent elements of a broader neuromorphic paradigm in which sensing, retention, and computation are not merely co-located but unified into a single adaptive element that updates its response based on its optical history~\cite{lowD_neuromorphic, wen20252d, guo2025recent}. vdW tunnel junctions accomplish this when operated as artificial optoelectronic synapses: the conductance state set by the history of optical pulses functions as the synaptic weight, each new optical input updates that weight locally, and the updated state is read out electrically without erasing it. The ReS$_2$/hBN/CuInP$_2$S$_6$ ferroelectric synapse discussed above is a complete demonstration of this paradigm in a tunneling architecture, emulating long-term plasticity, optical potentiation, and spike rate-dependent plasticity within a single junction, with artificial neural network simulations achieving 91\% classification accuracy~\cite{optoelectronic_synapse}.\\

Light-controlled tunnel junctions thus constitute a foundational building block for next-generation machine-vision and neuromorphic hardware. Co-locating sensing, memory, and computation within the same nanoscale junction through charge trapping, ferroelectric polarization switching, or band-alignment-tunable analog weighting enables adaptive, energy-efficient systems in which the traditional boundaries between detector, processor, and memory are dissolved~\cite{electricOptical_tunneling_review, wang2024multidimensional, liao2022bioinspired}.

\section{New opportunities with tunnel junctions}

This section highlights emerging opportunities enabled by recent advances in quantum materials and how these developments extend the use of tunneling for probing and engineering quantum states. In particular, ingredients such as quantum geometry, moiré superlattices, and momentum- and twist-resolved control can be naturally integrated into optically driven tunneling architectures, expanding both their spectroscopic reach and their functional capabilities. From a fundamental perspective, these developments allow tunneling currents to access aspects of electronic structure that are difficult to probe in equilibrium, including excited-state Berry curvature distributions, quantum metric contributions, and interaction-renormalized Floquet states. At the same time, they open pathways toward new functionalities, ranging from symmetry- and quantum geometry-sensitive tunneling to programmable, frequency-resolved control of quantum transport.

\subsection{Revealing quantum geometric quantities}

Quantum geometry describes the intrinsic geometric structure of Bloch wavefunctions in momentum space, characterized by the Berry curvature and the quantum metric --- capturing, respectively, the local phase structure of the wavefunction and the distance between neighboring quantum states in Hilbert space~\cite{xiao2010berry, ma2021topology, ma2023photocurrent}. These quantities are properties of the eigenstates themselves, independent of band energies, and are therefore difficult to access directly using conventional spectroscopic or linear-transport measurements. Their signatures have emerged across a broad range of phenomena: the anomalous Hall effect and orbital magnetism~\cite{xiao2010berry}, flat-band superconductivity and excitonic instabilities driven by quantum metric~\cite{torma2022superconductivity}, and the nonlinear Hall effect from Berry curvature dipoles~\cite{ma2019observation,tang2026bistable} and quantum metric dipoles~\cite{gao2023quantum}. Their optical fingerprints have been identified through nonlinear photocurrent responses: the circular photogalvanic effect encodes Berry curvature via helicity-dependent injection currents~\cite{xu2018electrically}, the shift current encodes the shift vector via linearly polarized bulk photovoltaic responses~\cite{ma2021topology, gao2025quantum}, and can be connected to Riemannian 
curvature of the quantum state manifold~\cite{ahn2022riemannian}.\\

\begin{figure}
\centering
\includegraphics[width=6.5in]{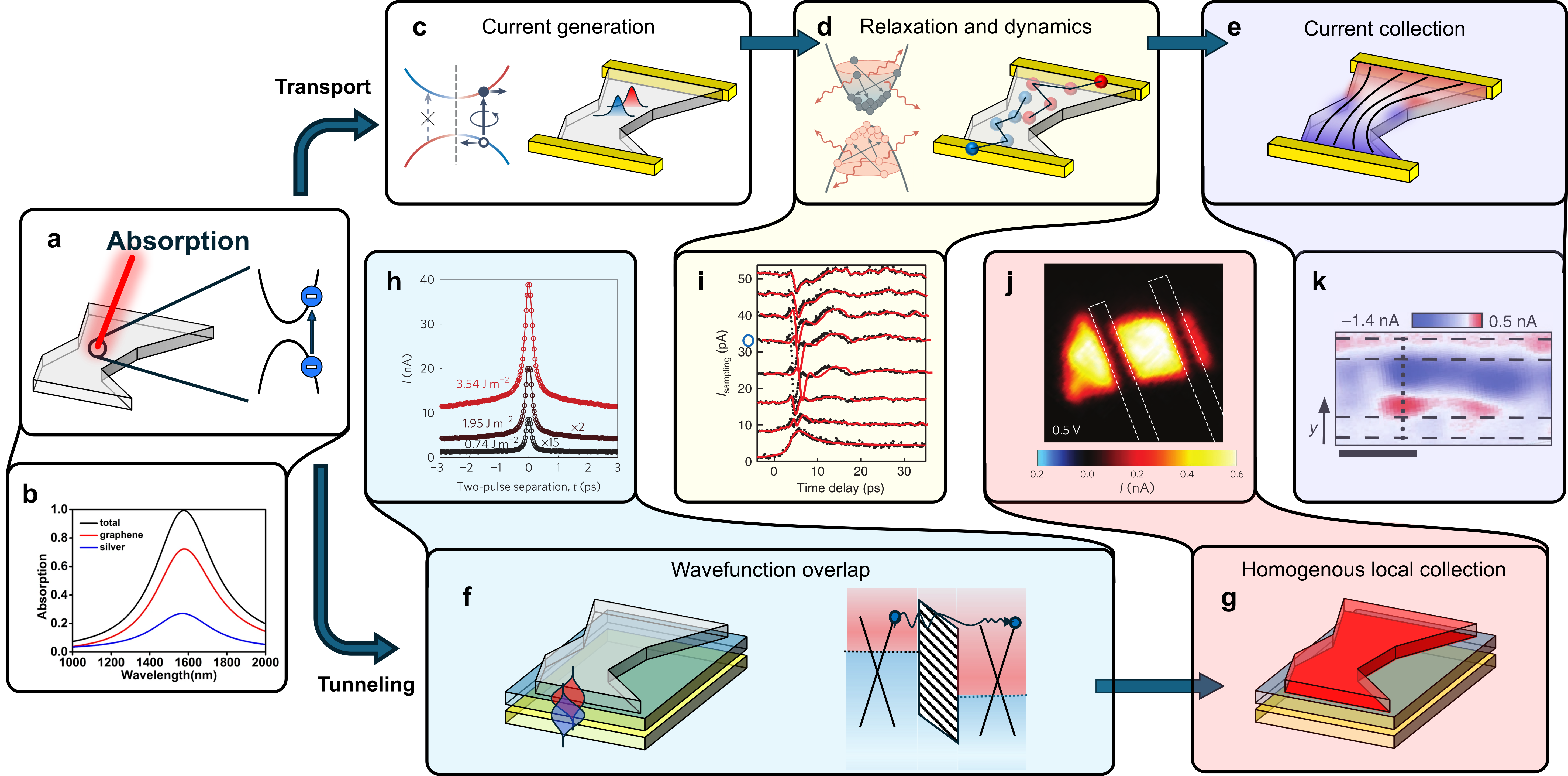}
\caption{
\textbf{Diffusive lateral photocurrent transport versus tunneling: the underlying physics in these two scenarios.}
\textbf{(a,b)} Light absorption creates photocarriers in a shared initial stage (graphene-based absorption example in \textbf{b}). Adapted from Ref.~\cite{graphene_absorption}.
\textbf{(c--e)} In-plane photocurrent readout proceeds through current generation, relaxation/scattering, and current collection, so the measured signal can include thermal or thermoelectric backgrounds and spatial inhomogeneity. Adapted from Ref.~\cite{ma2023photocurrent}.
\textbf{(f,g)} In vertical junctions, photocurrent is extracted locally through interlayer wavefunction overlap, reducing sensitivity to flake geometry and lateral current spreading. Adapted from Refs.~\cite{ma2023photocurrent}.
\textbf{(h,i)} Representative pump--probe traces for tunneling photocurrent and lateral photocurrent, respectively, showing sub-100-fs and \(\sim\)1-ps components in graphene/hBN/graphene tunneling devices, and a fast \(\sim\)4-ps response together with a slower \(\sim\)130-ps decay in lateral graphene photocurrent readout. Adapted from Refs.~\cite{Gr/BN/Gr_Ma,inplane_pc_time}.
\textbf{(j,k)} Representative spatial maps for tunneling photocurrent and lateral photocurrent, respectively, showing more uniform local collection in vertical junctions and position-dependent photocurrent, including sign-changing response, in lateral transport devices. Adapted from Refs.~\cite{Gr/BN/Gr_Ma,inplane_pc_time}.
These comparisons illustrate why tunneling-based readout can provide a more direct route to intrinsic quantum-geometric response.
}
\label{fig7}
\end{figure}

Existing transport and photocurrent experiments probing quantum geometry typically measure laterally averaged, in-plane responses that smear out the underlying quantum-geometric information. Between the moment of photon absorption and the creation of an excited electron with specific wavefunction-geometric character [Fig.~\ref{fig7}(a--c)] and the moment a signal is collected at contacts, there is an extended post-generation pipeline: carrier--carrier scattering, carrier--phonon relaxation, lateral diffusion, thermoelectric gradient formation, and current routing determined by the sample and contact geometry [Fig.~\ref{fig7}(d,e)]. Each step can dilute, convolute, and partially erase the intrinsic quantum-geometric imprint on the photoexcited distribution. Experimentally, this appears as spatially nonuniform photocurrent maps with hotspots and sign reversals governed by the sample and contact geometry rather than by the underlying photoexcited state [Fig.~\ref{fig7}(k)]. It also appears in pump--probe traces whose temporal dynamics typically span from a few to hundreds of picoseconds, reflecting carrier relaxation and thermalization rather than the generation event itself [Fig.~\ref{fig7}(i)]~\cite{tielrooij2015generation, inplane_pc_time}. Photocurrent nanoscopy improves spatial resolution but still relies on lateral current collection~\cite{ma2023photocurrent, SNOM_photocurrent_twist}. Polarization-resolved photoemission has recently provided access to the full quantum geometric tensor~\cite{kang2025measurements}, though its extreme surface sensitivity makes it less compatible with encapsulated or gated device geometries, and the deflection of emitted photoelectrons makes it less compatible with applied magnetic fields. It therefore remains challenging to detect angle- or momentum-resolved quantum geometry in a gate-tunable device.\\

Vertical tunneling offers a pathway that can reduce several of these complications [Fig.~\ref{fig7}(f,g)]. The tunneling process can more directly reflect the initial electronic state and its dynamics. For instance, sub-picosecond pump--probe measurements in graphene/hBN/graphene junctions show that tunneling can outcompete ultrafast intralayer carrier--carrier scattering [Fig.~\ref{fig7}(h)]~\cite{Gr/BN/Gr_Ma}. In addition, photocurrent is extracted perpendicularly through the barrier at the site of generation, rather than drifting laterally across the sample, reducing the influence of the sample and contact geometry and suppressing thermoelectric gradients that accumulate along lateral paths. This is manifested experimentally as a spatially uniform tunneling photocurrent map across the active overlap region [Fig.~\ref{fig7}(j)]~\cite{Gr/BN/Gr_Ma}. More fundamentally, the tunneling matrix element is governed by wavefunction overlap across the barrier, and this overlap depends on spin texture, orbital character, and geometric properties of the Bloch states, including Berry-curvature and quantum-metric contributions, in a way that goes beyond what band energies or the total density of states alone determine. In vdW heterostructures, atomically sharp interfaces and approximate conservation of in-plane crystal momentum make tunneling particularly sensitive to these wavefunction characters, because interfacial disorder that would otherwise average out wavefunction texture is suppressed. The tunneling photocurrent can then provide an electrically readable signature of the symmetry character generated at the moment of optical excitation~\cite{liang2026interferroic, inbar2023quantum, birkbeck2025quantum}.\\

One can therefore envision using light-controlled tunnel junctions to access quantum geometric information as follows. Optical excitation creates real carriers with defined momentum distributions and orbital character in the emitter layer, while vertical tunneling probes their coupling into collector states through a matrix element that retains the phase and texture of the underlying Bloch wavefunctions. By tuning photon energy and polarization, electrostatic gating, bias, and relative crystallographic alignment, one can isolate geometry-sensitive selection rules and track how the tunneling response follows changes in wavefunction texture. In realistic devices, the measured photocurrent integrates over in-plane momenta $\mathbf{k}_\parallel$ for which the collector provides states within the relevant energy window, and disorder broadens this further. Yet if the collector Fermi surface is anisotropic or its symmetry is reduced \cite{ehlen2018direct, zhao2015anisotropic, tang2026bistable, tang2024dual}, different momentum sectors are weighted unequally \cite{guo2025anisotropic, srivastava2021resonant, WS2/Te, yang2025bi2o2se}, and polarization-dependent signals that encode Berry curvature or quantum metric contributions may survive this integration. Momentum-conserving tunneling in aligned vdW stacks provides a natural implementation of this idea, because the junction itself acts as a momentum filter that helps preserve geometric information from being averaged out.\\

It is useful to distinguish two complementary light-driven limits that can both couple tunneling to quantum geometry. In the population picture emphasized above, illumination prepares real carriers in the existing Bloch bands, producing a nonequilibrium distribution whose momentum, spin, valley, and orbital selectivity is set by optical matrix elements and subsequent relaxation. The tunnel junction then acts as a wavefunction-sensitive detector of these photo-prepared states through momentum- and symmetry-selective overlap.\\

A distinct regime may arise under sufficiently strong and coherent periodic driving, where the electronic states are light-dressed and the emitter may be more naturally described within a Floquet framework~\cite{wang2013observation, mahmood2016selective, sentef2015theory, mciver2020light}. In that case, tunneling could proceed through the Floquet sideband structure rather than simply tunneling out of photo-populated equilibrium bands~\cite{gr_hBN_gr_floquet_theory, Anomalous_Floquet_tunneling_theory}. Hence, the quantum-geometry-sensitive selection rules discussed above, could be extended to tunneling involving Floquet states, providing a possible route to access quantum geometric information under periodic driving.\\

\subsection{Quantum twist microscope, tunnel junction, and light}

\begin{figure}
\centering
\includegraphics[width=6in]{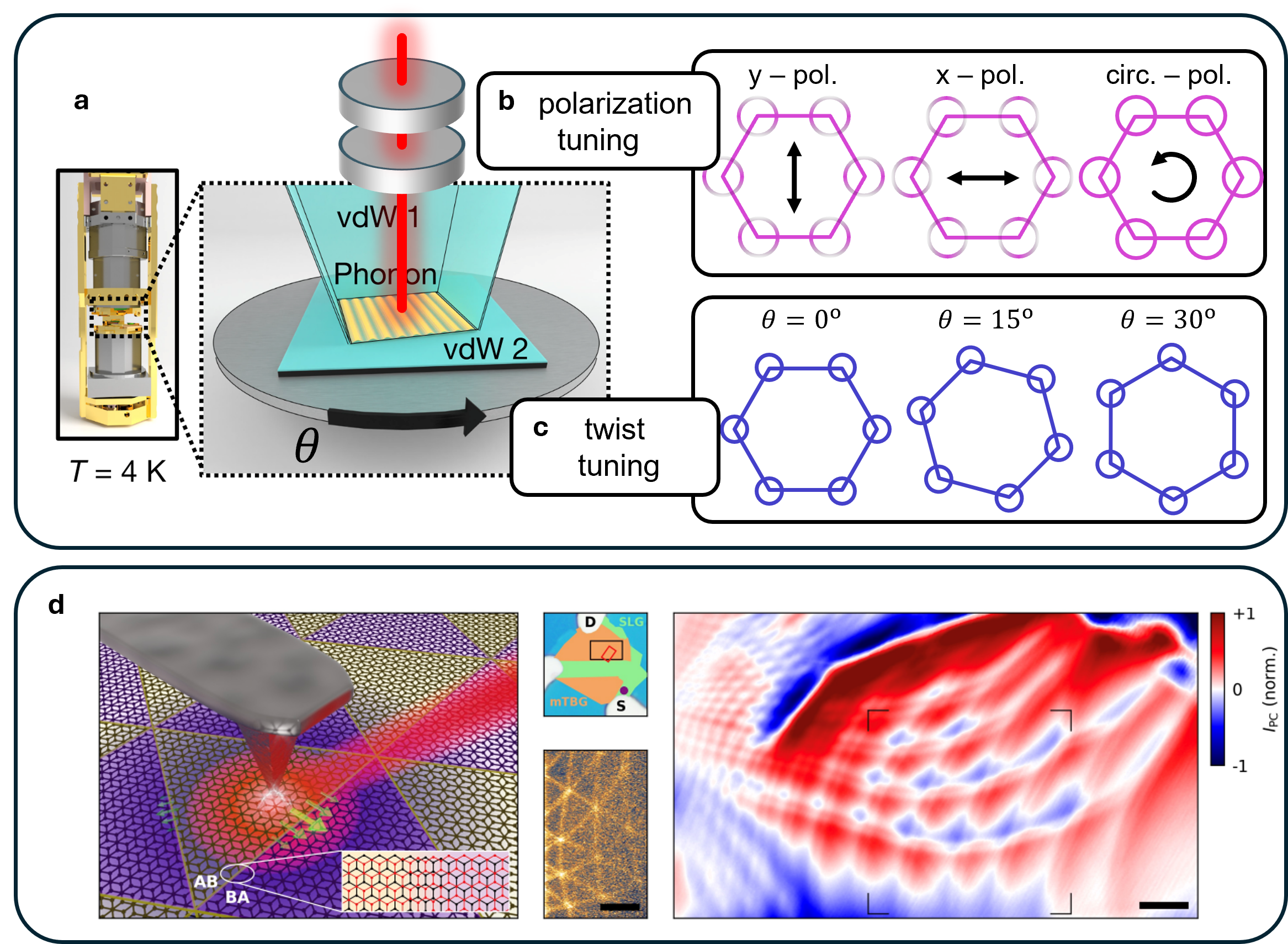}

\caption{\textbf{Twist- and spatially resolved photo-assisted tunneling for local momentum-selective imaging.}
\textbf{(a)} Concept of a light-integrated quantum twist microscope (QTM): the current of a vdW tip--sample junction is measured under continuous twist-angle tuning $\theta$ and optical excitation, extending QTM tunneling spectroscopy to photo-assisted operation. Adapted from Ref.~\cite{inbar2023quantum}.
\textbf{(b)} Polarization-resolved excitation ($x/y$ linear, circular) selectively populates different momentum sectors according to the optical selection rules; together with momentum-dependent wavefunction overlap in the tunneling process, this leads to different tunneling currents.
\textbf{(c)} Twist tuning rotates the relative momentum sectors, controlling interlayer momentum matching and wavefunction overlap.
\textbf{(d)} Near-field photocurrent nanoscopy in moir\'e graphene shows nanoscale, sign-changing local photoresponse tied to domain-wall and moir\'e structure. Adapted from Ref.~\cite{SNOM_photocurrent_twist}.
These elements show how twist control, polarization control, and local photocurrent imaging can be combined to probe momentum-selective quantum-material textures.
}

\label{fig8}
\end{figure}

Light-controlled tunnel junctions gain an additional control parameter when the twist angle between emitter and collector is made continuously tunable. This is precisely the capability introduced by the quantum twist microscope (QTM), which forms a clean, electrically contacted vdW interface between a stack mounted on a scanning tip and another on a substrate, while rotating their relative crystallographic angle $\theta$ in situ during measurement [Fig.~\ref{fig8}(a)]~\cite{inbar2023quantum, birkbeck2025quantum, xiao2024theory, xiao2025interacting, peri2024probing, klein2026imaging}. In the weakly coupled tunneling regime --- realized by inserting an insulating spacer or when out-of-plane momentum decay naturally confines transfer to evanescent overlap --- varying $\theta$ continuously shifts the alignment of the two Brillouin zones, selectively enhancing specific $k_\parallel$ channels. Sharp elastic resonances emerge at commensurate angles, and finite-bias spectra resolve inelastic features from phonon-assisted tunneling. By recording the tunneling current and differential conductance as functions of $\theta$, bias, and electrostatic gating, the QTM thus performs twist-resolved electrical spectroscopy of interlayer coupling, momentum-dependent hybridization, electron-phonon and electron-electron interactions~\cite{inbar2023quantum, birkbeck2025quantum, lee2025revealing}. In the strongly coupled (hybridized) regime, where the layers are brought into direct contact, $\theta$ instead engineers moiré band reconstruction, giving access to correlated and topological phases as a continuous function of twist angle~\cite{nuckolls2024microscopic, cao2018unconventional, cao2018correlated, choi2021correlation, xu2020correlated, tang2024chip}.\\

Adding light to this framework introduces a qualitatively new dimension. Polarization-dependent optical matrix elements weight different orbital, valley, and symmetry sectors of the emitter's photoexcited distribution, turning polarization into an independent spectroscopic axis complementary to bias and twist [Fig.~\ref{fig8}(b)]~\cite{gemotry_spotlight_M, ma2023photocurrent}. Meanwhile, varying $\theta$ rotates the relative Brillouin zones and therefore continuously scans which $k_\parallel$ sectors are geometrically allowed to participate in tunneling [Fig.~\ref{fig8}(c)]. Twist and polarization thus act as complementary handles on the same underlying wavefunction textures including orbital character, spin, and valley, and together provide access to geometry-sensitive responses that are difficult to isolate with either optical measurements or equilibrium tunneling alone~\cite{inbar2023quantum, birkbeck2025quantum, AFMTJ_review}.\\

The QTM platform is broadly applicable to a wide range of tunneling heterostructures. For example, junctions such as graphene/spacer/graphene, TMD/spacer/TMD, or TMD/spacer/ferromagnet enable twist-resolved tunneling under photoexcitation to probe momentum-selected band structures, optical selection rules, quantum geometry, and nonequilibrium carrier distributions. More complex configurations can incorporate moir\'e or correlated systems, for instance, magic-angle twisted bilayer graphene (MATBG)~\cite{cao2018correlated, cao2018unconventional}, twisted MoTe$_2$~\cite{cai2023signatures, anderson2023programming}, or rhombohedral graphene~\cite{han2024correlated, winterer2024ferroelectric}, across a tunneling barrier, where twist, bias, and optical excitation provide access to many-body states and correlation-induced renormalization of quantum geometry.\\

A complementary regime arises when no insulating spacer is present and interlayer hybridization is allowed. Here, twist can generate moiré reconstruction and strongly modify the low-energy electronic structure, so optical driving and photo-assisted tunneling can be used to probe interaction-driven phases~\cite{gr_hBN_gr_floquet_theory, Anomalous_Floquet_tunneling_theory}. This provides a bridge between momentum-selective tunneling spectroscopy and moiré-engineered quantum matter within a single twist-tunable platform. Because the QTM is a local probe, moreover, the same measurement can be repeated at different real-space positions, enabling systematic comparison across regions with varying strain, disorder, domain structure, or moiré potential landscape.\\

Looking forward, the combination of twist control and light-assisted tunneling provides a flexible framework for nonequilibrium spectroscopy: twist selects the dominant momentum-transfer channels, polarization selects optical selection rules, and bias/gating set the participating energy window. Together, these knobs extend twist-resolved measurements beyond equilibrium tunneling spectra toward time-dependent, light-controlled interlayer transport.

\subsection{Local photoresponse imaging with tunnel junctions}

A distinctive feature of vertical tunneling readout is that it can be made highly local in real space: the photocurrent is generated only where photoexcitation creates carriers that can be extracted across the barrier. This locality allows tunneling-based photocurrent to serve as an electrical probe of spatial inhomogeneity, because position-dependent absorption, relaxation, strain, disorder, and symmetry breaking are reflected directly in the measured photo-assisted tunneling response --- a sensitivity that is challenging for lateral photocurrent collection to match, since lateral collection integrates over extended diffusion paths that wash out local variations.\\

A common implementation is to scan a focused laser spot across a large-area tunnel junction and record the photocurrent as a function of position, constructing real-space maps of optoelectronic response. The diffraction limit constrains the beam spot to roughly half the optical wavelength, giving spatial resolution on the order of a few hundred nanometers for visible light. Such maps can be polarization-resolved and can reveal mesoscopic features that are often averaged out in global in-plane transport, including local strain fields, domain boundaries, and variations in barrier transparency.\\

To reach nanometer-scale resolution, the optical excitation itself must be confined beyond the diffraction limit. One route is to combine tunneling devices with scattering-type scanning near-field optical microscopy (s-SNOM), where the near-field tip concentrates the optical field into a nanoscale hotspot [Fig.~\ref{fig8}(b)]. The resulting photoexcited carriers are then read out electrically through the local vertical tunneling current, enabling nanoscale photocurrent maps that are sensitive to local trapping, exciton transport pathways, and spatially varying junction properties. Such near-field photocurrent approaches provide a device-integrated path toward sub-diffraction imaging of optoelectronic landscapes in vdW heterostructures~\cite{snom_photocurrent, snom_plasmon}.\\

A complementary strategy achieves nanoscale spatial resolution by confining the tunneling rather than the light: in photocurrent tunneling microscopy of twisted bilayer WS$_2$, broad optical illumination of an STM junction produced real-space images of photoexcited moiré excitons with sub-nanometer resolution, revealing spatially alternating photocurrent polarity within a single moiré unit cell that encodes the charge-transfer character of the in-plane moiré exciton~\cite{STM_photocurrent}. Together, these two strategies --- confining the light versus confining the tunneling --- define complementary limits of spatially resolved photo-assisted tunneling microscopy. Applying the light-confinement strategy to a vdW tunnel junction would preserve the sub-diffraction spatial resolution of near-field optics while replacing the mechanically constrained STM geometry with a clean, gate-tunable readout architecture.\\

An exciting direction to apply such local tunneling-based imaging is visualization of ferroelectric domains in vdW heterostructures, including both sliding and moiré ferroelectrics~\cite{zhang2025emerging, fei2018ferroelectric, zheng2020unconventional}. The ferroelectric order in these systems has been characterized by piezoresponse force microscopy~\cite{yasuda2021stacking}, Kelvin probe force microscopy~\cite{vizner2021interfacial}, and electrostatic force microscopy~\cite{woods2021charge}, as well as by transport through an adjacent graphene sensor~\cite{fei2018ferroelectric, yasuda2021stacking}. More recent work has shown that sliding ferroelectrics can also be incorporated directly into a tunnel junction with graphene electrodes, where polarization reversal modulates tunneling electroresistance~\cite{wang2026resonantFE, vareskic2025gate, gao2024tunnel}. This suggests that near-field optical probes combined with vertical tunneling readout could provide a direct way to image ferroelectric domains in an integrated, gate-tunable vdW junction. In such a geometry, the ferroelectric polarization resides within the tunnel barrier itself, the s-SNOM tip localizes the optically induced enhancement of the tunneling current to the nanometer scale, and the tunneling current serves as the electrical readout.\\

The same platform may also be extended from static domain imaging to ferroelectric switching dynamics. On top of the local optical probe described above, one can add a phase-controlled terahertz pump to drive polarization switching~\cite{mankowsky2017ultrafast}, while tracking the tunneling current as a function of pump--probe delay to access intrinsic switching dynamics with spatial resolution across ferroelectric domains in a gate-tunable vdW tunnel junction. Such measurements could clarify the microscopic switching pathway, including how polarization reversal nucleates, propagates, and evolves across the ferroelectric texture within the tunnel barrier~\cite{yasuda2024ultrafast}.

\subsection{New architectures for computing and sensing}

Photo-assisted tunnel junctions, particularly when implemented in low-dimensional and vdW material platforms, have already demonstrated promising potential for sensing and computing applications, including highly sensitive photodetectors with photon-energy-locked response~\cite{MoS2/BN/Gr_UV, ahmad2023progress}, electrically reconfigurable spectrometers~\cite{MoS2/WSe2_mini_spec}, in-sensor convolutional processors~\cite{PdSe2/MoTe2_processing}, and optoelectronic synaptic architectures~\cite{optoelectronic_synapse}. In their established form, these devices exploit the discrete energy exchange between electrons and photons to enable frequency-selective transport, rectification, and detection. However, these applications have largely relied on energy selectivity alone, treating the photon field as a scalar drive and the tunneling process as an energetically gated transport channel.\\

Incorporating quantum geometry, symmetry-resolved wavefunction structure, and twist- and momentum-selective tunneling expands the design space for computing and sensing architectures. In quantum materials with broken inversion symmetry, strong spin--orbit coupling, or valley-contrasting selection rules~\cite{TMDs_Mak, TMDs_DiXiao, MoS2_Mak}, the tunneling matrix elements depend explicitly on the polarization state of the driving field, so the junction current itself encodes polarization information and enables electrically readable polarization-sensitive detection without external optics; polarization thereby becomes an independent information channel enabling multiplexed sensing, chiral light detection, and symmetry-selective signal processing. Simultaneously, the incorporation of engineered band structures, van Hove singularities, and moiré minibands substantially enhances the intrinsic photon-energy sensitivity of the junction: small changes in photon energy can produce large changes in tunneling probability through resonant coupling to photon-dressed states, which could enable compact frequency-programmable spectrometers and, for computing, continuous analog control parameters for frequency-multiplexed or in-memory signal processing. Because the tunneling process responds to the instantaneous electromagnetic field, optical or terahertz driving further extends these capabilities into the time domain, enabling femtosecond-scale modulation of tunneling currents that is far beyond the reach of conventional electronic gating and that opens new pathways for time-resolved sensing of transient fields and nonequilibrium states.\\

These expanded capabilities motivate a system-level viewpoint that connects photo-assisted tunnel junction device physics to optical information processing. A useful abstraction is to treat a reconfigurable optoelectronic device as an encoder that converts unknown, high-dimensional optical inputs into a compact electrical response code, followed by a decoder that maps that code into the desired physical quantities or task-relevant outputs~\cite{geometric_optical_sensing_review}, as illustrated schematically in Fig.~\ref{fig9}(a). In this framework, the encoder is the photo-assisted tunnel junction pixel, and reconfigurability is the central resource: by changing gate voltages, interlayer bias, barrier asymmetry, and related electrical knobs, the same hardware implements a family of distinct transfer functions, each emphasizing different photon-assisted pathways by reshaping band alignment, coupling strength, symmetry selectivity, and momentum-matching conditions. Collecting responses under a small sequence of programmed operating states produces a compact electrical ``fingerprint'' that jointly encodes spectral, temporal, and polarization information into a compressed latent variable space. The decoder, implemented in a post-sensor processor, then operates on this small response code rather than a full high-dimensional data space, reducing data movement and downstream compute while maintaining performance for the target task~\cite{cui2025spectral}.\\

This multimodal optical information processing capability offers an efficient, compact, and low-latency route to edge artificial intelligence (AI) perception because it extracts complementary optical cues while minimizing optics overhead and readout bandwidth~\cite{wang2024multidimensional}. Many real scenes are underdetermined in intensity alone, so adding spectral and polarization contrast can disambiguate materials and surfaces, improve prediction robustness, and reveal features invisible in conventional imaging. Spectral information contributes chemical specificity and material fingerprints central to environmental monitoring, precision agriculture, and biomedical analysis where subtle spectral signatures correlate with composition and physiology~\cite{hyperspectral_imaging_review}. Polarization adds an orthogonal contrast channel in both linear and circular forms: linear polarization probes surface texture, orientation, and scattering anisotropy; circular polarization is directly linked to molecular and structural chirality, enabling sensitivity to chiral media and molecules through helicity-dependent tunneling responses~\cite{polarimetric_imaging_review}. This is especially compelling for biomedical sensing, where polarization-resolved imaging provides structural contrast across applications spanning dermatology, endoscopy guidance, and broader clinical assessment~\cite{polarisation_optics_for_bio_review}. In conventional systems, adding these modalities multiplies optical components, detectors, and readout channels. The encoder--decoder approach addresses this bottleneck by performing task-aware compression at the front end: electrically programmed photo-assisted tunneling measurements preserve the information most relevant to the inference goal while reducing downstream data movement and compute, which is an advantage for autonomous platforms, distributed sensors, and portable diagnostics where bandwidth and energy budgets are tightly constrained~\cite{wang2024multidimensional}.\\

From this perspective, multimodal photo-assisted tunneling sensing is not an incremental extension of frequency-selective detection. It is a path toward compact, programmable sensing and computing front ends where intrinsic energy selectivity and ultrafast temporal response are combined with symmetry-resolved wavefunction sensitivity to polarization and twist- or momentum-filtered selectivity tied to alignment and mode content. Electrical tuning supplies the codebook for generating informative response fingerprints, and decoding translates those compact codes into physical parameters or directly actionable outputs, reducing downstream data movement and hardware overhead while preserving task-relevant information.

\subsection{From single devices to system-level architectures}

While most experimental realizations of photo-assisted tunnel junctions to date have focused on individual devices or isolated junctions, an important next step is to elevate this platform from the single-device level to system-level architectures. The high degree of tunability, nonlinearity, and frequency selectivity already demonstrated at the device level makes individual junctions natural candidates for modular building blocks that can be replicated and interconnected~\cite{electricOptical_tunneling_review}: arrays configured in parallel or cascaded geometries could enable collective functionalities such as spatially resolved sensing, frequency-multiplexed signal processing, and multilayer analog computation, shifting the role of each junction from that of an isolated detector to that of a programmable element within a larger functional network. Realizing this transition requires addressing scalability, interconnectivity, and co-integration with complementary photonic, sensing, and electronic components.\\

\begin{figure}
\centering
\includegraphics[width=6in]{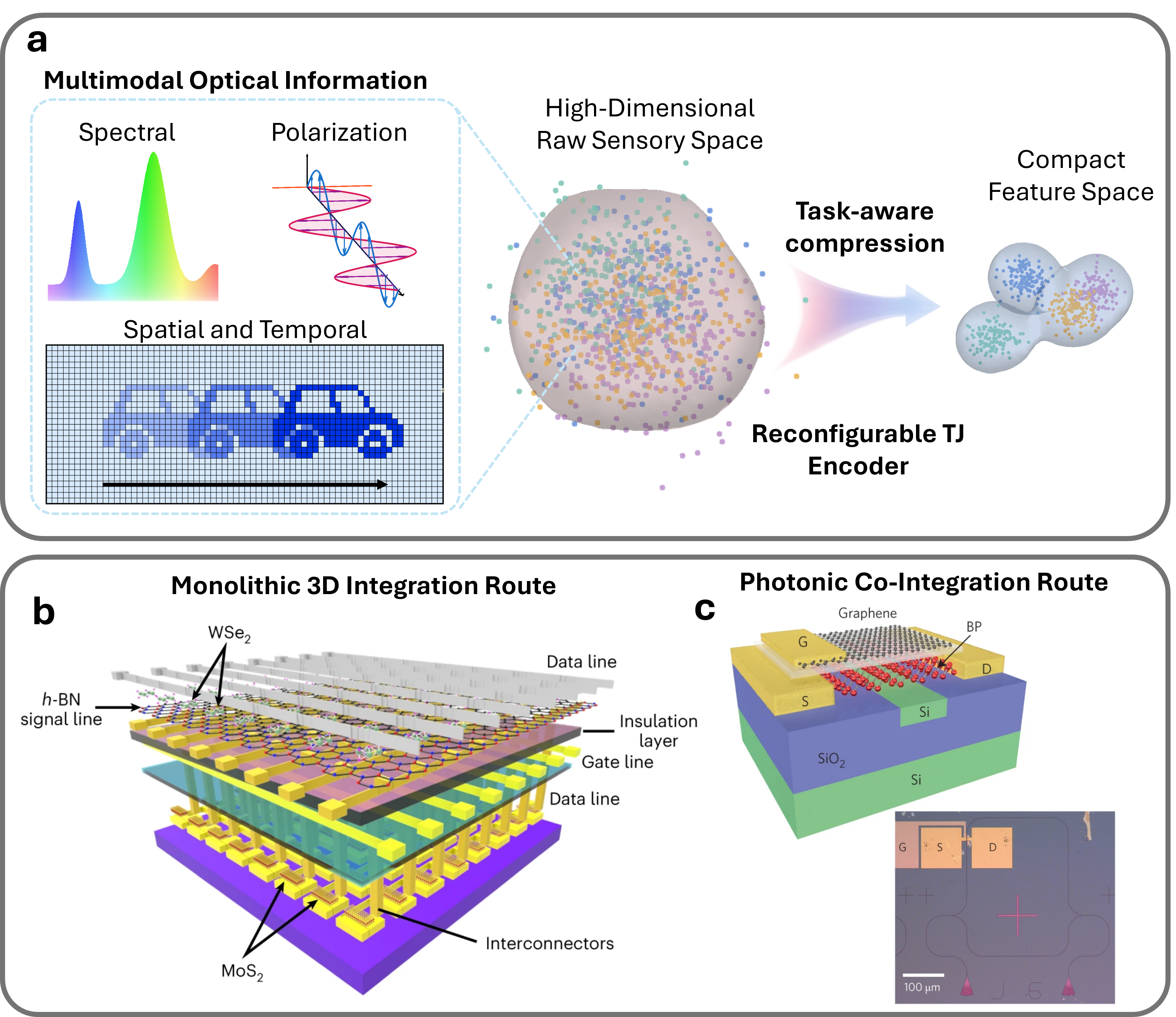}

\caption{\textbf{From programmable multimodal encoding to scalable system-level architectures.}
\textbf{(a)} Conceptual schematic of programmable multimodal encoding and task-aware compression. Multimodal optical inputs, including spectral, polarization, and spatiotemporal information, are mapped by a reconfigurable photo-assisted tunneling front end from a high-dimensional raw sensory space into a compact and more separable feature space, illustrating how physical encoding can reduce input redundancy and simplify downstream decoding or inference.
\textbf{(b)} Representative monolithic three-dimensional (M3D) integration architecture, illustrating a possible hardware route toward dense co-integration of sensing, memory, and computing functions. Shown is an M3D-integrated AI processor comprising WSe$_2$/hBN-based memristors and MoS$_2$-based transistors, where the drain electrode line of the transistors is connected to the bottom electrode of the memristors. Adapted from Ref. \cite{kang2023monolithic}.
\textbf{(c)} Representative on-chip photonic co-integration platform, illustrating a second route toward hybrid electro-optical system architectures. Shown are a three-dimensional device schematic featuring a few-layer graphene top gate and an optical microscope image of a complete device, in which a black phosphorus field-effect transistor is integrated into one arm of a Mach–Zehnder interferometer circuit. Adapted from Ref. \cite{youngblood2015waveguide}.
}
\label{fig9}
\end{figure}

Advancing toward such scalable architectures will depend critically on materials processing and device fabrication approaches that enable monolithic 3D integration with silicon-based or non-silicon-based computing technologies~\cite{3D_integration_2D_review}. The key engineering challenges are well-defined: non-uniformity in tunnel barrier thickness, interface quality, and contact resistance must be controlled to prevent response variation from degrading the benefits of front-end compression; surface planarization is essential because local topography introduced during multilayer transfer can induce strain and amplify device-to-device variation; and thermal management becomes a first-order constraint in dense 3D stacks, motivating co-design of heat-spreading layers and careful placement of temperature-sensitive sensing elements relative to high-power electronic blocks. A concrete demonstration of how these constraints can be navigated is provided by monolithic 3D-integrated AI processors combining WSe$_2$/hBN-based memristors with MoS$_2$ transistors, where direct electrode connection between the transistor drain and memristor bottom contact enables dense co-integration of sensing, memory, and computing functions within a single stack [Fig.~\ref{fig9}(b)]~\cite{kang2023monolithic}. These are challenges shared broadly by 2D-material integration efforts and are being actively addressed through advances in clean-room transfer, atomic-layer-deposition encapsulation, and heterogeneous bonding processes~\cite{3D_integration_2D_review}.\\

A particularly promising direction for system-level integration is the incorporation of photo-assisted tunneling devices into on-chip photonic platforms. Integrated optics provides a scalable and low-loss means of delivering, routing, and manipulating optical or terahertz fields with high spatial precision. A proof of concept for this co-integration has already been demonstrated: graphene/hBN/graphene tunnel junctions coupled directly to Si$_3$N$_4$ waveguides enable compact metal--insulator--metal tunneling field-effect transistors locally excited by guided optical modes~\cite{Gr/BN/Gr_Si3N4_waveguides}, eliminating the need for free-space optical alignment. A complementary route integrates 2D material photodetectors directly into interferometric photonic circuits: embedding a black phosphorus field-effect transistor into one arm of a Mach--Zehnder interferometer on-chip achieves high responsivity with low dark current while preserving full photonic circuit functionality [Fig.~\ref{fig9}(c)]~\cite{youngblood2015waveguide}, establishing a clear pathway toward dense electro-optical integration.\\

The combination of photo-assisted tunneling with on-chip photonics offers several system-level advantages. First, it enables precise spatiotemporal control of optical excitation, allowing different tunnel junctions on the same chip to be addressed independently or synchronously. Second, photonic waveguides naturally support frequency- and polarization-multiplexed operation, which aligns directly with the intrinsic frequency and polarization sensitivity of photo-assisted tunneling. Third, from an architectural perspective, integrated photonics also enables hierarchical system design. Optical signals can be generated, modulated, routed, and combined on-chip before interacting with photo-assisted tunneling junctions, which then perform nonlinear transduction, mixing, or detection. Electrical outputs from these junctions can be further processed using conventional electronics or fed into subsequent tunneling stages, enabling hybrid electro-optical systems that leverage the strengths of both domains. Such architectures are particularly attractive for applications requiring high bandwidth, low latency, and compact form factors.\\

\noindent\textbf{Acknowledgments:} 
Q.M. acknowledges support from the U.S. Department of Energy under Award No.~DE-SC0026332 and the Office of Naval Research under Grant No.~N00014-24-1-2102 during the development of this review article. P.J.H. and Q.M. acknowledge the Air Force Office of Scientific Research under Grant No.~FA9550-21-1-0319. PJH acknowledges support by the Office of Naval Research (grant no. N000142412440), the MIT/Microsystems Technology Laboratories Samsung Semiconductor Research Fund, the Gordon and Betty Moore Foundation’s EPiQS Initiative through grant GBMF9463, and the Ramon Areces Foundation. Y.C.L. acknowledges the partial support from Texas A$\&$M University and the National Science Foundation, under award no. DMR-2425545.\\

\noindent\textbf{Author contributions:} All authors wrote the manuscript together.\\

\noindent\textbf{Competing interests:} The authors declare that they have no competing interests.

\newpage

\end{document}